\documentclass[12pt]{article}
\usepackage{amssymb} 
\usepackage{graphicx}
\usepackage{amsmath}

\def\e{{\rm e}}

\def\d{\partial}

\newcommand{\be}{\begin{equation}}
\newcommand{\ee}{\end{equation}}
\newcommand{\bea}{\begin{eqnarray}}
\newcommand{\eea}{\end{eqnarray}}
\newcommand{\bg}{\begin{gather}}
\newcommand{\eg}{\end{gather}}
\newcommand{\bseq}{\begin{subequations}}
\newcommand{\eseq}{\end{subequations}}
\newcommand{\tg}{\mathop{\rm tg}\nolimits}
\newcommand{\ctg}{\mathop{\rm ctg}\nolimits}

\newcommand{\sh}{\mathop{\rm sh}\nolimits}
\renewcommand{\ln}{\mathop{\rm ln}\nolimits}

\renewcommand{\Im}{\mathop{\rm Im}\nolimits}

\newcommand{\bpm}{\begin{pmatrix}}
\newcommand{\epm}{\end{pmatrix}}
\newcommand{\arcsh}{\mathop{\rm arcsh}\nolimits}

\newcommand{\arcth}{\mathop{\rm arcth}\nolimits}
\makeatletter

\def\compoundrel#1\over#2{\mathpalette\compoundreL{{#1}\over{#2}}}
\def\compoundreL#1#2{\compoundREL#1#2}
\def\compoundREL#1#2\over#3{\mathrel
         {\vcenter{\hbox{$\m@th\buildrel{#1#2}\over{#1#3}$}}}}
\makeatother

\begin{document}
\begin{flushright}
CERN-PH-TH/2007-065
\end{flushright}

\vspace*{1cm}
\begin{center}
{\Large\bf  On the over-barrier reflection in quantum mechanics with
multiple degrees of freedom}

\vspace*{0.6cm}
D.G.~Levkov$^{a}$\footnote{levkov@ms2.inr.ac.ru},
A.G.~Panin$^{a,b}$\footnote{panin@ms2.inr.ac.ru},
S.M.~Sibiryakov$^{c,a}$\footnote{Sergey.Sibiryakov@cern.ch, 
~sibir@ms2.inr.ac.ru}\\
$^a${\small
Institute for Nuclear Research of the Russian Academy of Sciences,}\\  
{\small 60th October Anniversary prospect 7a, Moscow 117312, Russia.}\\
$^b${\small Moscow Institute of Physics and Technology,} \\
{\small Institutskii per. 9, Dolgoprudny 141700, Moscow Region, Russia.}\\
$^c${\small Theory Group, Physics Department, CERN, CH-1211 Geneva 23,
  Switzerland.}
\end{center}
\vspace{0.5cm}

\begin{abstract}
  We present an analytic example of two dimensional quantum mechanical
  system, where the exponential suppression of the 
  probability of over--barrier reflection
  changes non-monotonically with energy.
The suppression is minimal at certain ``optimal'' 
energies where reflection occurs with exponentially larger probability
than at other energies.
\end{abstract}

\section{Introduction}
\label{sec:introduction}
Tunneling and over--barrier reflection are the 
characteristic non--perturbative phenomena in quantum
mechanics. They typically occur
with exponentially small probabilities,
\begin{equation}
\label{P}
{\cal P} \propto \mathrm{e}^{-F/\hbar}\;,
\end{equation}
where $F$ is the suppression exponent; still, the above phenomena
are indispensable in understanding a wide variety of physical situations,
from the generation of baryon number asymmetry in the early
Universe~\cite{Kuzmin:1985mm} to chemical reactions~\cite{Miller} and
atom ionization processes~\cite{Perelomov}. 

During the last decades extensive investigations of tunneling
processes in systems with many degrees of freedom have been
performed~\cite{Miller,Heller:1981,Wilkinson:Takada,Miller:2001,chaotic,
Bonini:1999cn,Bezrukov:2003,Drew,Mechanism-Takahashi,Stokes,Stokes-Shudo}.
These studies revealed a rich variety of features of multidimensional
tunneling which are in striking contrast to  the properties of
one--dimensional tunneling and over--barrier reflection.
In particular, the following phenomenon has been observed:
the probability of tunneling may depend non-monotonically
on the total energy of the system and exhibit resonance-like peaks.
One can envisage three
physically different mechanisms of this phenomenon.
The first mechanism, present already in
one-dimensional case, is tunneling via creation of a metastable
state. In this 
case the tunneling probability at the maximum of the 
resonance is exponentially
higher than at other energies. On the other hand, the resonance width
$\Delta E$ 
is exponentially suppressed; so, after averaging with an 
energy distribution of a finite width the effect of the resonance is
washed out in the semiclassical limit $\hbar\to 0$.
The second possible mechanism of non-monotonic behavior of ${\cal P}(E)$
is quantum interference
\cite{chaotic,Stokes-Shudo} (see also \cite{Levkov:2007ce}). In this
case the peak value of the tunneling probability is only by a
factor of order one higher than the average value, while the
width of the resonance scales as $\Delta E\propto\hbar$. Again,   
the resonances become indiscernible in the semiclassical limit.
In both these cases the
resonances can be attributed to the subleading semiclassical
corrections, i.e. non-monotonic behavior of the
pre-exponential factor omitted in Eq.~(\ref{P}). 
The third
possibility is that the suppression exponent $F(E)$ is
non-monotonic. In this case the existence of the ``resonances'' is the
leading semiclassical effect: the optimal tunneling probability
at the maximum of the resonance is
exponentially higher than the probability at other energies. At the same time  
the resonance width scales as\footnote{This follows from the
  representation 
$${\cal P}(E)\propto \exp{\left(-\frac{F(E_o)}{\hbar}
-\frac{F''(E_o)(E-E_o)^2}{2\hbar}\right)}$$
of the tunneling probability in the vicinity of the maximum.
}
$\Delta E\propto\sqrt\hbar$. 
This last possibility of ``optimal tunneling'' is definitely
of interest; yet, it did not receive much attention in literature. We are
aware of only a few works mentioning non--monotonic dependence of the
suppression exponent on energy \cite{Ioselevich:1986,Voloshin:1993dk,
Levkov:2007ce}. It is worthwhile studying 
this phenomenon in detail; this can provide
a new insight into the dynamics of multidimensional tunneling. 

In this paper we consider the process of over--barrier reflection in a
simple model with two degrees of freedom. Our setup is interesting in
two  respects. First, the model under study is essentially
non--linear and the variables cannot be separated; still, 
over--barrier reflections in this model can be described
analytically within the semiclassical framework. Thus, this model can
serve as an analytic laboratory for the study of multidimensional
tunneling. Second, the
suppression exponent $F$ of the reflection process behaves
non--monotonically as the total energy $E$ changes. We demonstrate
that the function $F(E)$ possesses a number of local 
minima $E = E_o$, where reflection is optimal. We stress
that  the process we study is {\it exponentially}
preferable at ``optimal'' energies as compared to other energies.

\begin{figure}
\centerline{\includegraphics[angle=-90,width=\textwidth]{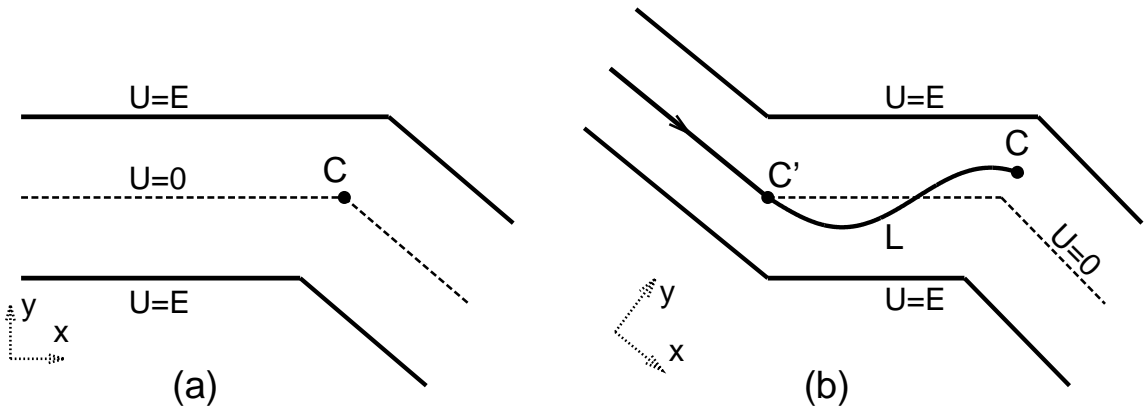}}
\caption{The equipotential contour $U=E$
  for the waveguides with (a) one and (b) two sharp turns. An example
  of classical trajectory is shown in  the case
  (b).}
\label{fig:11}
\end{figure}

Our model describes the motion of a quantum particle in 
the two dimensional harmonic waveguide (see
Refs.~\cite{Bonini:1999cn,Drew,Levkov:2007ce} for similar models). The
Hamiltonian is 
\begin{equation}
\nonumber
H = \frac{p_x^2}{2 m} + \frac{p_y^2}{2 m} + \frac{m\omega^2}{2}
w^2(x,y)\;,
\end{equation}
where $x$, $y$ are the Cartesian coordinates and  $m$ is the mass of
the particle. The function $U = m\omega^2 w^2/2$ represents the waveguide
potential in two dimensions: a particle with small energy is bound
to move along the  line $w(x,y)\approx 0$. We 
do not introduce a potential barrier across the waveguide and consider
the case when the line $w=0$ stretches all the way from $x\to -\infty$
to $x\to +\infty$. We also assume that the function $w(x,\,y)$ is
linear in the initial asymptotic region,
\begin{equation}
\nonumber
w(x,y) \to y \qquad \mbox{as}\qquad x\to -\infty\;.
\end{equation}  
In the present paper we consider two particular cases of
the function $w(x,y)$ describing waveguides with one and two sharp
turns\footnote{The explicit expressions
  for the waveguide functions $w(x,y)$ will be presented in the
  subsequent sections.}, see Fig.~\ref{fig:11}. 

The motion of the particle at $x\to -\infty$ is a superposition of 
free translatory 
motion in $x$ direction and oscillations of frequency 
$\omega$ along $y$ coordinate;
the state of such a particle is fully 
characterized by two quantum numbers, the total energy $E$ and
$y$--oscillator excitation number $N$. 
The particle sent into the waveguide from the 
asymptotic region ${x\to-\infty}$ with given $E$, $N$ may
either continue to move towards $x\to +\infty$, or
reflect back into the region $x\to -\infty$. We are interested in the 
probability 
${\cal P}(E,N)$ of reflection. 

Let us discuss reflections at the classical level. [Note that the classical
counterpart of $N$ is the energy of transverse oscillations.]
Consider first the waveguide with one sharp turn 
(Fig.~\ref{fig:11}a). One observes that the outcome of the classical
evolution, i.e. whether or not the particle 
reflects from the turn, depends not only on the total energy $E$, but on
other dynamical quantities as well. In particular, the
direction of the momentum of the particle in the vicinity of the turn
(point
$C$ on the graph) is important. This means that the entire
dynamics in the waveguide should be taken into account
in order to determine the possibility of classical reflection.
This is in sharp contrast with the situation in one--dimensional
case, where 
reflection from the potential barrier (or transition through it)
is ensured by the value of the conserved energy of the particle. 

Now, consider the waveguide with two turns.
The model is characterized by the angles of the turns and the distance $L$
between them
(see Fig.~\ref{fig:11}b). 
Suppose the particle starts moving classically from $x\to -\infty$
with $N=0$ along the valley $w=0$.
Then, the transverse oscillations get
excited only after the particle crosses the first turn, point $C'$ on
the plot, so that at
the time of arrival to the second turn (point $C$) approximately 
$\omega \tau/2\pi$ oscillations are made, where $\tau \sim L\sqrt{m/2E}$
is the time of motion between the two turns. The state
of the particle (coordinates and momenta) at which it
comes across the second turn depends
{\it periodically} on the phase of transverse oscillations
$\omega\tau$. Hence, one expects that 
the regime of motion of the classical particle can change from
transmission to reflection and back as the energy grows ($\tau$
decreases); the energies where it
happens can be roughly estimated as
\be
\label{Eo}
E_n\sim\frac{m\omega^2L^2}{2(2\pi n)^2}\;.
\ee
We will see that this is indeed the case for the waveguides
with certain angles of the turns.

At some values of $E$, $N$ the reflection process cannot proceed
classically. Then, at the quantum mechanical level its probability is
exponentially suppressed, $F(E,N)> 0$.
It is natural to call such a process ``over--barrier
reflection''\footnote{By this term we want to emphasize that the
  process is classically forbidden.
Recall, however, that there is
no actual potential barrier across the waveguide in our setup.}. 
The central quantity to be studied below is the suppression
exponent $F(E,N)$ of this process. The above
discussion suggests that $F(E,N)$, being determined by the entire
dynamics in the waveguide, may be a highly non--trivial function. For
the particular case of the waveguide with alternating regimes
of classical reflections and transmissions $F$ should 
oscillate: $F=0$  at the energies where the classical reflections are
allowed, and $F>0$ at the energies where the reflections are
classically 
forbidden. One can expect that the similar oscillatory behavior of the
suppression exponent persists for other two--turn models as well. 
Now, instead of reaching zero, $F$ may possess 
a number of local positive minima  implying that 
the reflection at the ``optimal'' energies is still a tunneling process.

Let us emphasize the difference of the
``optimal tunneling'' from quantum interference and resonance
phenomena in our two--turn model. 
The interference of the de Broglie waves reflected from
the two turns can, in principle, lead to oscillations in the reflection
probability ${\cal P}(E)$. One can estimate the positions of the
interference peaks by equating the De Broglie wavelength of the
particle to an integer fraction of the distance between the turns,
$2\pi\hbar/\sqrt{2mE}\sim L/n$. This yields the energies of the
interference peaks,
\[
E^{int}_n\sim \frac{(2\pi n)^2\hbar^2}{2mL^2}\;.
\]  
This formula is completely different from Eq.~(\ref{Eo}) for the peaks
due to ``optimal tunneling''. In particular, the 
distance between the adjacent inteference peaks, 
\[
\Delta E^{int}\sim \frac{2\pi\hbar}{L}\sqrt{\frac{2E}{m}} \;,
\]
scales proportional to $\hbar$. Thus, these peaks should be averaged
over in the semiclassical limit. Besides, the amplitude of the
interference peaks
is at most of order one and does not affect the suppression
exponent. Indeed, the exponential increase of
the scattering amplitude can arise due to quantum interference only in
the presence of a resonant state with 
exponentially long life--time. This state should be supported
somewhere in between the turns and should be classically
stable. In Sec.~\ref{sec:classical-evolution} we show that such states
are absent in our system. One concludes that 
the peak--like structure of the probability ${\cal P}(E)$ 
of ``optimal tunneling'' is caused by  completely different physical
reasons as compared to the case of resonance scattering in quantum 
theory.

It is worth noting that the phenomenon of ``optimal tunneling'' has an
important implementation in field theory. 
Recently it was argued~\cite{Levkov:2004tf}
(see also Ref.~\cite{Voloshin:1993dk}) that the probability of
tunneling induced by particle collisions~\cite{induced,Rubakov:1992ec}  reaches its
maximum at a certain ``optimal'' energy and stays 
constant\footnote{As opposed to the quantum mechanical case, the
  tunneling probability does not decrease at energies higher than the
  ``optimal'' one. This is due to the possibility, specific to the
  field theoretical setup, to emit the excess of energy into a few
  hard particles, so that tunneling effectively occurs at the
  ``optimal'' energy.} 
at higher energies. 
This result, if generic, provides the answer to the long--standing
question~\cite{Ringwald:1989ee}
about the high--energy behavior of the probability of
collision--induced nonperturbative transitions in field theory.  
The quantum mechanical model presented here supports the generic
nature of the phenomenon of ``optimal tunneling''; the simplicity of
our model enables one to get an intuitive insight into the nature of 
this phenomenon. 

The paper is organized as follows. In
Sec.~\ref{sec:semiclassical-method} we review the semiclassical
method of complex trajectories, which is exploited in the rest
of the paper. Reflections in the waveguides with one and two turns are
considered in Secs.~\ref{sec1turn} and~\ref{sec:two-turn-model}
respectively. 
We discuss our results in Sec.~\ref{sec:discussion}.
In appendix we analyze the validity of some assumptions 
made in the main body of the paper. 

\section{The semiclassical method}
\label{sec:semiclassical-method}
We start by describing the semiclassical 
method\footnote{Note that the method has been confirmed
  by the explicit comparison with the exact quantum mechanical results
  in Refs.~\cite{Bonini:1999cn,Bezrukov:2003,Levkov:2007ce}; specifically, the recent
  check~\cite{Levkov:2007ce} deals with the case when the dependence of the
  suppression exponent on energy is not monotonic.} of complex
trajectories which will be used in the study of 
over--barrier reflections. We concentrate on the derivation of the
formula for the suppression exponent $F(E,N)$ (see
Refs.~\cite{Miller,Bonini:1999cn,Bezrukov:2003} for the details of the method and
Ref.~\cite{Rubakov:1992ec} for the 
field theory formulation). In what follows we  use the system of units 
$$
\hbar = m = \omega = 1\;,
$$
where the Hamiltonian takes the form,
\begin{equation}
\label{H}
H = \frac12 \left( p_x^2 + p_y^2 + w^2(x,\,y) \right)\;.
\end{equation}

One starts with the  amplitude of reflection into the state with definite 
coordinates ${x_f<0\;, y_f}$,
\begin{equation}
\label{A}
{\cal A} = \langle x_f,\,y_f | \mathrm{e}^{-i\hat{H}(t_f - t_i)}
| E,\, N\rangle\;.
\end{equation}
Here $|E,\, N\rangle$ is the initial state of the particle moving 
in the asymptotic region $x_i \to -\infty$ with fixed translatory momentum
$p_0 = \sqrt{2(E-N)}$ and the oscillator excitation number $N$. Semiclassically,
\begin{equation}
\label{BBi}
\langle x_i,\, y_i | E,\, N\rangle = \mathrm{e}^{ip_0 x_i} \cos\left(
\int_{\sqrt{2N}}^{y_i} p_y(y')dy' + \pi/4\right)\;,
\end{equation}
where $x_i$, $y_i$ denote initial coordinates,
\begin{equation}
\label{py}
p_y(y') = \sqrt{2N-y'^2}\;,
\end{equation}
and we omitted the pre-exponential factor which is irrelevant for our
purposes. 
Using Eq.~\eqref{BBi},  
one rewrites the amplitude \eqref{A} as a path integral,
\begin{equation}
\label{AA}
{\cal A} = \int dx_i dy_i \int [dx] [dy]
\Bigg|_{x_i,\,y_i}^{x_f,\,y_f}\, \mathrm{e}^{iS + ip_0 x_i} \cos\left( 
\int_{\sqrt{2N}}^{y_i} p_y(y') dy' + \pi/4\right)\;,
\end{equation}
where $S$ is the classical action of the model \eqref{H}.

In the semiclassical case the integral (\ref{AA}) 
is dominated by the (generically complex) saddle point.
Note that, as we continue the integrand in Eq.~\eqref{AA} into the plane
of complex coordinates, one of the exponents constituting the initial 
oscillator wave function grows, while the other becomes 
negligibly small. Within the validity of our approximation, 
we omit the decaying exponent by writing
\begin{equation}
\label{sub}
\cos\left(\int_{\sqrt{2N}}^{y_i} p_y(y') dy' + \pi/4\right) \to
\mathrm{exp}\left({i\int_{\sqrt{2N}}^{y_i} p_y(y') dy'}\right)\;,
\end{equation}
with the standard choice\footnote{\label{f1} 
The correct branch is fixed by 
drawing a cut between the oscillator turning points $y = \pm \sqrt{2N}$, 
and choosing $\mathrm{Im}\, p_y > 0$ at $y\in {\mathbb{R}}$, $y>\sqrt{2N}$,
see, e.g., Refs.~\cite{Elyutin2001}. 
}
of the branch of the square root in Eq.~(\ref{py}).

One proceeds by finding the saddle point for the integral \eqref{AA} with 
the substitution \eqref{sub}. 
Extremization with respect to $x(t)$, $y(t)$ leads to the classical equations
of motion,
\begin{equation}
\label{Tthetaeq}  \ddot{x} = - w w_x \;,\qquad\qquad
 \ddot{y}  = -w w_y\;.
\end{equation}
Differentiating with respect to $x_i \equiv x(t_i)$, 
$y_i \equiv y(t_i)$, one obtains,
\begin{equation}
\notag
\dot{x}_i = p_0 = \sqrt{2(E-N)}\;,\qquad\qquad
\dot{y}_i = p_y(y_i) = \sqrt{2N - y_i^2}\;.
\end{equation}
The latter equations are equivalent to fixing the total
energy $E$ and  initial 
oscillator energy $N$ of the complex trajectory,
\bseq
\label{Tthetabc-1}
\begin{align}
& E = \frac{1}{2} \dot{x}_i^2 + N\;,\\
& N = \frac{1}{2} \left( \dot{y}_i^2 +  y_i^2 \right)\;.
\end{align}
\eseq
Substituting the saddle--point configuration\footnote{For simplicity
  we assume that the saddle--point configuration is
  unique. Otherwise, one should take the saddle point corresponding to
  the weakest exponential suppression.} 
into Eq.~\eqref{AA}, 
one obtains the amplitude of the process with exponential accuracy,
\begin{equation}
\nonumber
{\cal A} \propto \mathrm{e}^{iS + iB(x_i,\,y_i)}\;,
\end{equation}
where the term
\begin{equation}
\label{Bi}
B(x_i,\,y_i) = p_0 x_i +  \int_{\sqrt{2N}}^{y_i} p_y(y') dy' 
\end{equation}
is the initial--state contribution. 
For the inclusive reflection probability one writes,
\begin{equation}
\nonumber
{\cal P} = \int dx_f dy_f\, |{\cal A}|^2  \propto 
\int dx_f dy_f \,\mathrm{e}^{
iS - iS^* + i B - i B^*}\;.
\end{equation}
The integral over the final states can also be evaluated by the saddle point
technique; extremization with respect to $x_f\equiv x(t_f)$, $y_f\equiv y(t_f)$
fixes the boundary 
conditions in the asymptotic future, 
\begin{equation}
\label{Tthetabc+}
 \Im\dot{x}_f = \Im x_f = 0\;,
\qquad\qquad \Im \dot{y}_f = \Im y_f = 0\;.
\end{equation}
In this way one obtains the expression \eqref{P} for the reflection 
probability, where the suppression exponent $F$ is given by the 
value of the functional 
\begin{equation}
\nonumber
F(E,\, N) = 2\Im S + 2 \Im B(x_i,\, y_i)
\end{equation}
evaluated on the saddle--point configuration --- a complex
trajectory satisfying the boundary value problem \eqref{Tthetaeq},
\eqref{Tthetabc-1}, \eqref{Tthetabc+}. 

The contribution $B(x_i,\, y_i)$  of
the initial state is simplified
after one uses the asymptotic form 
of the solution at $t\to -\infty$ ($x_i \to -\infty$),
\begin{equation}
\label{xas}
 x = p_0 t + x_0  \;, \qquad \qquad
 y = a\mathrm{e}^{-it} + \bar{a} \mathrm{e}^{it}\;.
\end{equation}
Equations \eqref{Tthetabc-1} guarantee  that the 
quantities $p_0 = \sqrt{2(E-N)}$ and $2a\bar{a} = N$ are real, 
since $E,\, N \in 
{\mathbb{R}}$. Therefore, one may introduce two real parameters 
$T$, $\theta$ as follows,
\begin{equation}
\label{Tthetabc-2}
 2\Im\, x_0 = -p_0T \;,\qquad\qquad
 \bar{a} = a^* \mathrm{e}^{T +\theta}\;.
\end{equation}
One finds for the initial term \eqref{Bi},
\begin{align}
\notag
2\Im\,B(x_i,\, y_i) &= \Im\, \left\{ 
2 p_0 x_i - 2N \mathrm{arccos}(y_i/\sqrt{2N}) + y_i \sqrt{2N - y_i^2}
\right\} \\\notag
&= -p_0^2 T-N (T+\theta) + \Im 
(y_i \dot{y}_i)\;,
\end{align}
and thus 
\begin{equation}
\label{F}
F = 2\Im\, \tilde{S} -ET - N\theta \;,
\end{equation}
where $\tilde{S}$ is the classical action of the system \eqref{H} 
integrated by parts,
\begin{equation}
\label{Sbyparts}
\tilde{S} = - \frac{1}{2}\int_{t_i}^{t_f} dt\,\left[x\ddot{x} + y\ddot{y} 
+ w^2(x,\,y) \right]\;.
\end{equation}

Let us comment on the physical meaning of the parameters $T$, $\theta$.
Consider two trajectories which are solutions to
the boundary value problem (\ref{Tthetaeq}), (\ref{Tthetabc-1}),
(\ref{Tthetabc+}) at neighbouring  values of $E$, $N$.
The differential of the quantity $2\Im\tilde{S}$ as one deforms one
trajectory into the other is
\begin{equation}
\nonumber
d\,(2\Im\tilde{S}) = d\Im(2S + x_i\dot{x}_i +y_i \dot{y_i}) = 
\Im(x_i d\dot{x}_i - \dot{x}_id x_i + y_i d\dot{y}_i - \dot{y}_i dy_i)
 = EdT + Nd\theta\;,
\end{equation}
where in the last equality 
we used the asymptotic form (\ref{xas}), (\ref{Tthetabc-2}) of the solution. 
Then, from Eq.~\eqref{F} one finds,
\begin{equation}
\label{dF}
dF(E,N) = -TdE - \theta dN\;.
\end{equation} 
Thus, the parameters $T$ and $\theta$ are (up to sign) the derivatives 
of the 
suppression exponent with respect to energy $E$ and initial oscillator 
excitation number $N$ respectively. 

Our final remark is that the boundary value problem 
\eqref{Tthetaeq}, \eqref{Tthetabc-1}, \eqref{Tthetabc+}
is invariant with respect to the trivial time translation symmetry,
\begin{equation}
\label{TimeTranslations}
t \to t + \delta t\;, \qquad \delta t\in \mathbb{R}\;,
\end{equation}
which can be fixed in any convenient way.

\section{The model with one turn}
\label{sec1turn}
\begin{figure}
\centerline{\includegraphics[angle=-90,width=0.7\textwidth]{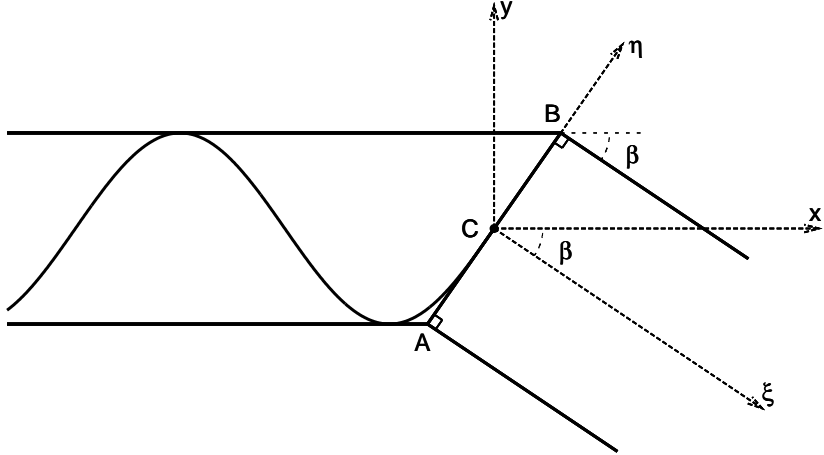}}
\caption{\label{fig1} The equipotential contour $w^2(x,y) = 2N$ for the 
waveguide \eqref{a1} and the 
trajectory of the critical solution with energy $N/\cos^2\beta$.}
\end{figure}

To warm up, we 
consider the simplest model, where the waveguide has one 
sharp turn,
\begin{equation}
\label{a0}
w = y\,\theta(-x+y\tg\beta)+\cos\beta\,
(x\sin\beta+y\cos\beta)\,\theta(x-y\tg\beta)\;.
\end{equation}
Here $\theta(x)$ is the step function. It is convenient to use the 
rotated coordinate system,
\begin{equation}
\nonumber
\left(\begin{array}{c}\xi \\ \eta\end{array}\right) = 
\left(\begin{array}{cc}\cos\beta & -\sin\beta \\ 
\sin\beta & \cos\beta \end{array}\right)
\left(\begin{array}{c}x\\ y\end{array}\right)\;.
\end{equation}
The waveguide function takes the form,
\begin{equation}
\label{a1}
w = \eta\cos\beta-\xi\sin\beta\;\theta(-\xi)\;.
\end{equation}
The equipotential contour $w^2(\xi,\,\eta)=\mathrm{const}$ 
is shown in Fig.~\ref{fig1}. 
One observes that the motion of the particle 
in two regions, $\xi<0$ and $\xi>0$, decomposes into the translatory 
motion and oscillations in the coordinates $x$, $y$ 
and $\xi$, $\eta$ respectively (see. Eqs.~(\ref{a0}) and (\ref{a1}));
the frequency of $\eta$--oscillations  
in the latter case is $\cos \beta$.

Due to the presence of the step function, 
the first derivatives of the potential \eqref{a1} are 
discontinuous\footnote{Note that the
  potential itself is continuous.} at $\xi=0$. Strictly speaking, the
semiclassical method is not applicable in this situation \cite{Elyutin2001}.
Thus, the formula \eqref{a1} should be regarded as 
an approximation to some waveguide function with smooth
turn. Generically the width of the smoothened turn is characterized by
a parameter $b$; the sharp--turn approximation (\ref{a1}) corresponds
to $b\to 0$. An example of smoothening is provided by the following
substitution in Eq.    
\eqref{a1},
\begin{equation}
\label{thetab}
\theta(\xi) \to \theta_b(\xi) = \frac{1}{1+\mathrm{e}^{-\xi/b}}\;.
\end{equation}
The semiclassical description can be used as long as the de
 Broglie wavelength of the particle is small compared to the linear
 size of the potential\footnote{Another semiclassical condition is
 that the energy is sufficient to excite a lot of oscillator levels,
 $E\gg1$. It is satisfied provided Eq. (\ref{qc}) holds.},
 $1/\sqrt{E}\ll b$.
We conclude that the sharp--turn and
semiclassical  approximations are valid simultaneously for 
 smooth waveguides with 
\begin{equation}\label{qc}
 1\gg b\gg 1/\sqrt{E}\;.
\end{equation}

An important property of the model \eqref{a1} is 
invariance of the
classical equations of motion \eqref{Tthetaeq} under the  
rescaling of the coordinates,
\begin{equation}
x\to \Lambda x\;, \qquad y\to \Lambda y\;. \label{ReflSymm}
\end{equation}
Using the transformation \eqref{ReflSymm}, one may express a
solution $x(t)$,  
$y(t)$ with energy $E$ in terms of the ``normalized'' one,
\begin{equation}
\nonumber
x = \tilde x \sqrt{E}\;, \qquad y = \tilde y\sqrt{E}\;, 
\end{equation}
where the solution $\tilde x(t)$, $\tilde y(t)$ has unit energy; its 
initial oscillator excitation number is 
$$
\nu=N/E\;.
$$
The suppression exponent \eqref{F} takes the form,
\begin{equation}
\label{F1}
F(E,\;N) = E f_{\beta}(\nu)\;,
\end{equation}
where $f_{\beta}(\nu)$ is 
the exponent for the ``normalized'' solution.
Substituting the expression \eqref{F1} into Eq.~\eqref{dF}, one obtains,
\begin{equation}
\label{f}
f_{\beta}(\nu) = -T - \theta\nu\;.
\end{equation}
We will exploit Eq. (\ref{f}) in the end of this
section. Now, we proceed to finding the ``normalized'' trajectories.

At certain initial data
$\nu>\nu_{\mathrm{cr}}$ the particle can reflect from the 
turn classically, so that
$$
f_{\beta}(\nu>\nu_{\mathrm{cr}}) = 0\;.
$$ 
Let us find the value of $\nu_{\mathrm{cr}}$.
In the region $\xi < 0$ the classical solution takes the form,
\bseq
\label{ClassAllTraj}
\begin{align}
&x(t) = p_0 t + x_0\;,\\
&y(t) = A_0 \sin(t + \varphi)\;.
\end{align}
\eseq
Having crossed the line $\xi=0$ (line $AB$ in 
Fig.~\ref{fig1}), the classical 
particle can never return back into the region $\xi<0$. 
Indeed, in this case it moves at $\xi>0$ with
constant momentum $p_\xi>0$. Thus, the particle can reflect classically 
only if its trajectory touches the line $\xi=0$. 
The potential of our model has ill--defined derivatives at $\xi=0$,
and the
fate of the particle moving along the line $AB$ depends
on the particular choice of the smoothening of the potential. In
appendix we consider the  
motion of the classical particle in the case when nonzero smoothening of
width $b$ is
switched on. For a class of smoothenings
we show that in the small vicinity ($\delta \xi \sim b$) of
any trajectory touching the line $\xi=0$ there exists some 
``smoothened'' trajectory, which reflects classically from the turn.
Consequently, below we associate the trajectories touching the line $\xi=0$
with the classical reflected solutions. 

One notices that the inclination of the trajectory \eqref{ClassAllTraj} is 
bounded from above
$$
\left| \frac{dy}{dx} \right| \leq \frac{A_0}{p_0}\;;
$$
therefore, the classical trajectory of the particle can touch the line
$\xi=0$, that is, $y/x = \mathrm{ctg}\, \beta$
only at 
\begin{equation}
\label{nucr1}
A_0/p_0 \geq \mathrm{ctg}\,\beta\;.
\end{equation}
From Eqs.~\eqref{nucr1}, \eqref{ClassAllTraj}, 
\eqref{Tthetabc-1} one extracts the condition for the particle to reflect 
classically from the turn,
\begin{equation}
\label{nucr}
\nu\geq \nu_{\mathrm{cr}} = \cos^2\beta\;.
\end{equation}
The critical classical solution at $\nu=\nu_{\mathrm{cr}}$ 
touches the line $\xi=0$ at $\eta=0$ (point $C$ in 
Fig.~\ref{fig1}), where 
its trajectory
\begin{align}
& x_{\mathrm{cr}}(t) = \sqrt{2} t \sin\beta\;,\label{Critical}\\
& y_{\mathrm{cr}}(t) = \sqrt{2} \sin t \cos{\beta}\;. \notag
\end{align}
has the largest inclination.

We now turn to the classically forbidden reflections at 
$\nu< \nu_{\mathrm{cr}}$, which are described by the boundary value
problem \eqref{Tthetaeq}, \eqref{Tthetabc-1}, \eqref{Tthetabc+}. 
One makes the following important observation. The waveguide 
function  \eqref{a1} has the form of two analytic functions glued 
together at $\xi=0$. Hence, the equations of motion \eqref{Tthetaeq} can
be continued analytically to the complex values of coordinates in two different
ways, starting from the regions $\xi<0$ and $\xi>0$ respectively. In
this way
one obtains two complex solutions, $\xi_-(t)$, $\eta_-(t)$ and $\xi_+(t)$,
$\eta_+(t)$. These solutions and their first derivatives should be
matched at some moment of time $t_1$, $\xi(t_1) = 0$.
[Note that the matching time $t_1$ does not need to be real.]
Below we conventionally refer to these solutions as the ones belonging to 
the regions $\xi<0$ and $\xi>0$. 

By the same reasoning as above we find that once the 
particle arrives into the region $\xi>0$, it never reflects back to
$\xi<0$, unless $p_\xi=0$. So, in the region $\xi>0$ one writes,
\bseq
\label{theta1+}
\begin{align}
&\xi_+(t) = 0\;,\\
& \eta_+(t) = \frac{\sqrt{2}}{\cos\beta} \sin(t\cos\beta 
+ \varphi_\eta)\;,
\end{align}
\eseq
where the ``normalization'' condition $E=1$ has been used explicitly. 
Due to the conditions in the asymptotic future, 
Eqs.~\eqref{Tthetabc+}, the parameter
$\varphi_\eta$ is real. 
We use the translational invariance 
\eqref{TimeTranslations} to set $\varphi_{\eta}=0$.
Note that we again associate the trajectory going along the line $\xi=0$
with the reflected one. 

The physical picture of over--barrier reflection that comes to mind
matches with the new mechanism of multidimensional tunneling proposed
recently in Refs.~\cite{Bezrukov:2003,Mechanism-Takahashi}. The
process proceeds in two steps. The first step, which is exponentially
suppressed, is formation of the periodic classical orbit
(\ref{theta1+}) oscillating along the line $\xi=0$. This orbit is
unstable. At the second step of the process the unstable orbit  
decays classically forming a trajectory going back to $x\to-\infty$ at
$t\to+\infty$. Clearly, the second step does not affect the
suppression exponent of the whole process, and we do not consider it
explicitly. In what follows we 
concentrate on the determination of the tunneling trajectory
describing the first step of the process.
 
One should find the solution at $\xi < 0$ and impose the boundary 
conditions \eqref{Tthetabc-1}. Note, however, 
that the energy of our solution is fixed already. As for the initial
oscillator excitation number $\nu$, it does not change during the evolution
in the region $\xi < 0$. Thus, one may fix it at the matching time $t = t_1$.
One writes,
$$
\nu = \frac12 (\dot{y}^2 + y^2) \Bigg|_{t = t_1} = \cos^2 \beta +
\sin^2\beta \sin^2(t_1 \cos\beta)\;.
$$
This complex equation allows one to express $t_1$ as
\begin{equation}
\label{sew1_1} 
\sin(t_1 \cos\beta) = -i\frac{\sqrt{\nu_{\mathrm{cr}} - \nu}}{\sin{\beta}}\;,
\end{equation}
where the choice of the sign is dictated by the condition in
footnote~\ref{f1}.
It is convenient to introduce notation 
$t_1 = iT_1$, $T_1\in \mathbb{R}$.

\begin{figure}
\centerline{\includegraphics[width=0.7\textwidth]{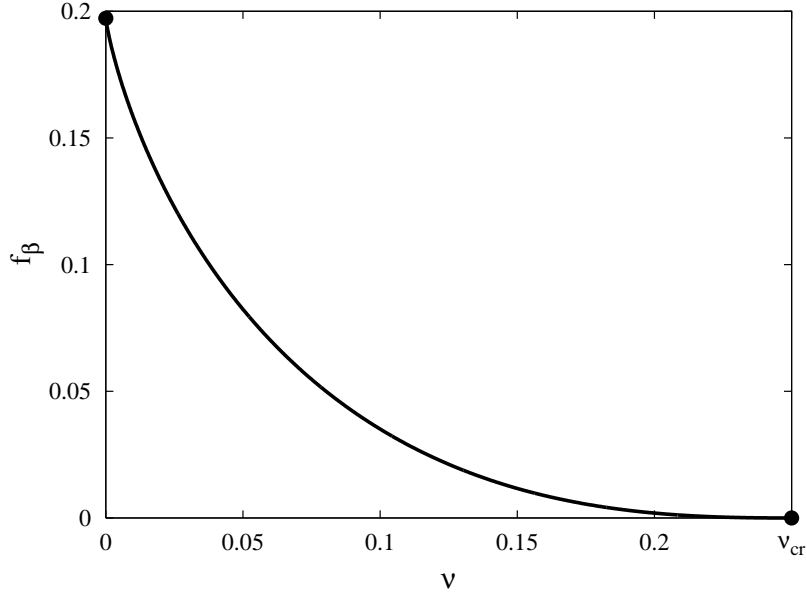}}
\caption{\label{fig4} The suppression exponent $f_{\beta}(\nu)$ for the 
waveguide \eqref{a1}; $\beta=\pi/3$. }
\end{figure}

In order to find the suppression exponent $f_{\beta}(\nu)$,
one  needs to 
evaluate the parameters $T(\nu)$, $\theta(\nu)$.  
At $\xi<0$ the solution has the form,
\bseq
\label{theta1-}
\begin{align}
& x_-(t) = p_0(t - iT/2) + x_0'\;,\\
& y_-(t) = a\mathrm{e}^{-it} + a^* \mathrm{e}^{T+\theta + it}\;,
\end{align}
\eseq
where the definitions \eqref{xas}, \eqref{Tthetabc-2} 
have been taken into account explicitly,
so that  $p_0,\, x_0' \in \mathbb{R}$. 
One evaluates $p_0$, $x_0'$, $a$, $T$, $\theta$ by matching the 
coordinates $x_{\pm}$, $y_{\pm}$ and their first derivatives $\dot{x}_{\pm}$,
$\dot{y}_{\pm}$ at $t = iT_1$; this yields
\begin{align}
x_0' = 0\;, &\qquad p_0 = \sqrt{2(1-\nu)}\;,
\qquad a = i\sqrt{\frac{\nu}{2}}\,\e^{-\frac{T+\theta}{2}}\;,\notag\\
\qquad T_1  - \frac{T}{2} &= 
-\sqrt{\frac{1 - \nu/\cos^2\beta}{1 - \nu}}\;,\qquad
\sh\bigg(T_1-\frac{T+\theta}{2}\bigg)=
-\frac{\sqrt{\cos^2\beta-\nu}}{\sin\beta\sqrt\nu}\notag
\end{align}
The last two equations, together with Eq.\eqref{f},
define the function $f_{\beta}(\nu)$,
\begin{equation}
\nonumber
f_{\beta}(\nu) = \frac{2}{\cos\beta}\left\{  
\arcsh{\frac{\sqrt{\nu_{\mathrm{cr}}-\nu}}{\sin\beta}-
\nu\cos\beta\arcsh\frac{\sqrt{\nu_{\mathrm{cr}}-\nu}}{\sin\beta\sqrt\nu}}
- \sqrt{(\nu_{\mathrm{cr}}-\nu)(1-\nu)}\right\}\;;
\end{equation}
this finction is plotted in Fig.~\ref{fig4}. 
One observes that at $\nu\to\nu_{\mathrm{cr}}$ the quantities 
$T_1,\;T,\;\theta,\;f_{\beta}$ tend to zero, and 
the complex trajectory tends to the classically allowed 
critical solution, cf. Eqs.~\eqref{Critical},
$$
 p_0 \to \sqrt{2} \sin{\beta}\;, \qquad a \to \frac{i}{\sqrt{2}} 
\cos{\beta}\;. 
$$
At $\nu= 0$ one has,
\begin{equation}
\label{f0}
 f_{\beta}(0) =  -2 + \frac{2}{\cos{\beta}} \arcth\, (\cos\beta)\;.
\end{equation}
To summarize, we obtained  
the suppression exponent for the
reflection of a particle in the simplest waveguide with one
sharp turn. 

\section{The model with two turns}
\label{sec:two-turn-model}
\subsection{Introducing the system}\label{sec:introducing-system}
In the model of the previous section the suppression exponent was
proportional to energy because of the coordinate rescaling 
symmetry \eqref{ReflSymm}.
Now, we are going to demonstrate that small 
violation of this symmetry  
results in highly non--trivial graph for $F(E)$.

\begin{figure}
\centerline{\includegraphics[angle=-90,width=\textwidth]{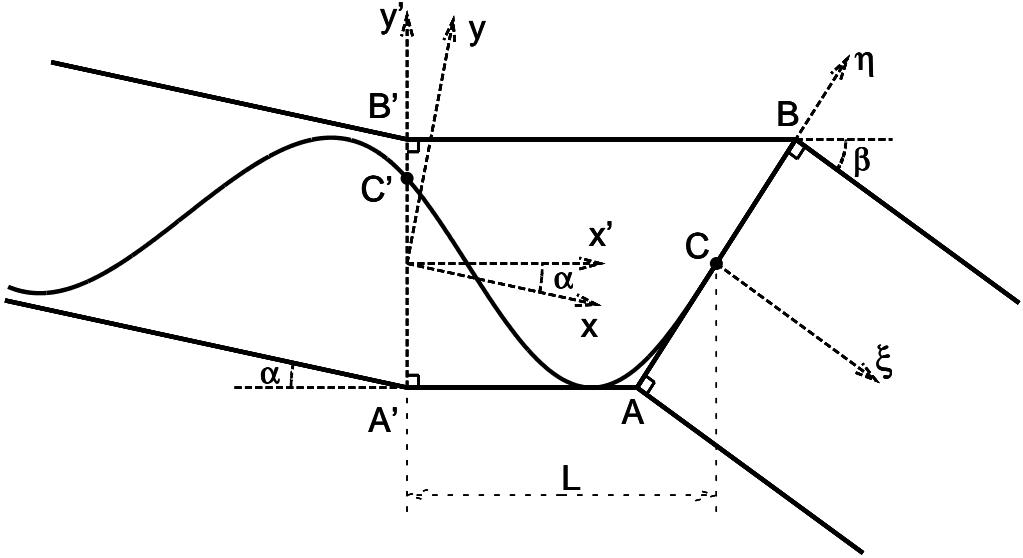}}
\caption{The equipotential contour $w^2(x,y) = 2N'$ 
for the waveguide 
\eqref{a2} and the trajectory 
of the critical solution with energy $N'/\cos^2\beta > E_B$. The matching
points $C$, $C'$ are shown by the thick black dots. }
\label{fig5}
\end{figure}
One introduces a second turn into the waveguide, see Fig.~\ref{fig5}. We
want to consider this turn as a small perturbation, so, we assume its
angle $\alpha$ to be smaller than $\beta$.
It is convenient to introduce two additional coordinate
systems, $x'$, $y'$ and $\xi$, $\eta$, bound to the central and
rightmost parts of the waveguide respectively. They are related to the
original coordinate system $x$, $y$ as follows,
\begin{equation}
\left(\begin{array}{c}x'\\ y'\end{array}\right) = \left(\begin{array}{cc}
\cos\alpha & \sin\alpha \\ -\sin\alpha & \cos\alpha\end{array}\right)
\left(\begin{array}{c} x\\ y\end{array}\right)\;, \;\;\;\;\;\;
\left( \begin{array}{c} \xi \\ \eta \end{array}\right) = 
\left( \begin{array}{cc} \cos\beta & -\sin\beta \\ \sin\beta & \cos\beta
\end{array}\right) \left( \begin{array}{c} x'-L \\ y'\end{array}\right)\;.
\label{xsysxieta}
\end{equation} 
Note that the origin of the coordinate system $\xi$, $\eta$ is 
shifted by the distance $L$. The waveguide function is
\begin{equation}
\label{a2}
w = \theta(-x')\theta(-\xi) y + \theta(-\xi)\theta(x') y' 
\cos \alpha  + \theta(\xi)
\eta \cos\alpha \cos\beta\;;
\end{equation}
it consists of three pieces glued 
together continuously at $x'=0$ and $\xi=0$ (lines $A'B'$ and 
$AB$ in Fig.~\ref{fig5} respectively). At $t\to -\infty$  the
particle comes flying from the asymptotic  
region $x'<0$, where $w=y$.
In the intermediate region $x'>0$, $\xi < 0$ the particle moves in the 
$x'$ direction oscillating  along the $y'$ coordinate with the frequency 
$\cos\alpha$. Finally, in the region $\xi>0$ its motion is 
free
in the coordinates $\xi$, $\eta$; the frequency of 
$\eta$--oscillations is $\cos\alpha \cos\beta$. 

The model \eqref{a2} no longer possesses the symmetry 
\eqref{ReflSymm}: rescaling of coordinates changes the 
length $L$ of the central part of the waveguide. In what follows it is
convenient to work in terms of the rescaled dynamical variables,
$$
\tilde{x} = x/L\;, \qquad \tilde{y} = y/L\;.
$$
In new terms the 
parameter $L$ disappears from the classical equations of motion,
entering the theory through the overall coefficient $L^2$ in front of
the action.
The initial--state
quantum numbers are also proportional to $L^2$, 
\be
\label{tilde1}
E = L^2 \tilde{E}\;,\qquad N = L^2 \tilde{N}\;.
\ee
Thus, the conditions (\ref{qc}) for the validity of the semiclassical
approximation are satisfied in the limit
\begin{equation}
\nonumber
L\to \infty\;, \qquad \tilde{E},\;\tilde{N} = \mathrm{fixed}\;. 
\end{equation}
The suppression exponent takes the form
\be
\label{tilde2}
F(E,N) = L^2 \tilde{F}(\tilde{E},\tilde{N})\;.
\ee
To simplify notations, we omit tildes over the rescaled quantities in
the rest of this section. Rescaling back to the physical units can be
easily performed in the final formulae by implementing Eqs. 
(\ref{tilde1}), (\ref{tilde2}).

\subsection{Classical evolution}\label{sec:classical-evolution}
Let us begin this subsection by demonstrating that there are no stable
classical solutions localized in the region between the turns. This is
important for the determination of the tunneling probability,
since such stable solutions  could lead to exponential
resonances in the tunneling amplitude. 
The argument proceeds as follows. Any trajectory which is localized in
the intermediate region should reflect from the line $AB$ infinitely
many times. Each reflection involves touching the unstable
orbit living at the line $AB$. This implies that the trajectory
itself is unstable.

\begin{figure}
\centerline{\includegraphics[angle=-90,width=0.7\textwidth]{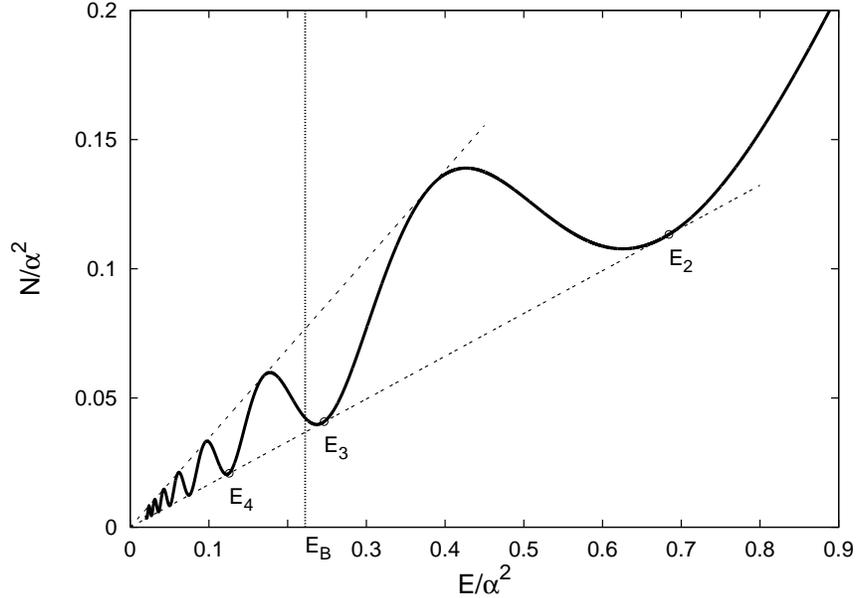}}
\caption{The boundary $N = N_{\mathrm{cr}}(E)$ of the classically
  allowed region at $E>E_B$ for the waveguide model 
\eqref{a2}; $\beta=\pi/3$, $\alpha = \pi/30$. 
The region of the classically allowed initial data lies above this
boundary. 
The empty circles correspond to the energies $E = E_{n}$, where the curve 
$N = N_{\mathrm{cr}}(E)$ touches its lower envelope 
$N = E\cos^2(\beta+\alpha)$.
}
\label{fig6} 
\end{figure}

We proceed by determining the region of initial data $E$, $N$, which
correspond to the classical reflections. [For brevity we will refer 
to this
region as the ``classically allowed region'', as opposed to the
``classically forbidden region''
where reflections occur only at the quantum mechanical level. We
stress that these are the regions in the plane of quantum numbers $E$,
$N$.] 
Let us search for the 
critical classical solutions which
correspond to the smallest initial oscillator number $N =
N_\mathrm{cr}(E)$ at 
given energy $E$. As in the previous section, one finds that 
the particle must 
get stuck at the line\footnote{We do not consider reflections from the line
  $A'B'$. They disappear at larger values of $N$ than reflections from the line
  $AB$ if $\alpha$ is small enough.} $AB$ for some 
time in order to reflect back. 
Let us first make an assumption inspired by the study of the one-turn
model that the critical solutions touch the line $AB$ at their maximum
inclination point (point $C$ in Fig.~\ref{fig5}). We will see shortly
that this
 is true only at
energies above a certain value $E_B$, see Eq.~(\ref{EB}). 
Still, the analysis based on the above assumption enables one to catch the
qualitative features of the critical line  $N=N_\mathrm{cr}(E)$. Besides,
the analysis is considerably simplified in this case;  
we postpone the accurate study until the end of this
subsection. Keeping in mind the above remarks, one writes for
the solution in the intermediate region,
\bseq
\label{Critical2}
\begin{align}
&x_{\mathrm{cr}}'(t) = t\sqrt{2E} \sin\beta + 1\;,\\ 
& y_{\mathrm{cr}}'(t) = \sqrt{2E}\; \frac{\cos \beta}{\cos\alpha}
\sin (t\cos\alpha) \;.
\end{align}
\eseq
Before entering the intermediate region, the particle crosses the
line $A'B'$ (point $C'$ in Fig.~\ref{fig5}). The initial oscillator
number $N$ is most conveniently calculated at the moment 
\begin{equation}
\nonumber
t = t_0 \equiv -\frac{1}{\sqrt{2E}\sin\beta}
\end{equation}
of crossing.
Using the relations \eqref{xsysxieta} one obtains,
\begin{equation}
\label{p0cr}
\dot{x}_{\mathrm{cr}}(t_0) = \sqrt{2E}\left[\sin\beta \cos\alpha -
\cos\beta \sin\alpha\cos\left(\frac{\cos\alpha}{\sqrt{2E}\sin\beta}
\right)\right]\;,
\end{equation}
and thus
\begin{equation}
\label{Ncr}
N_{\mathrm{cr}}(E) = E - \frac{1}{2}\dot{x}_{\mathrm{cr}}^2 (t_0) = 
E - E\left[\sin\beta \cos\alpha - \cos\beta \sin\alpha \cos\left(
\frac{\cos\alpha}{\sqrt{2E}\sin\beta}\right)\right]^2\;,~~~~E>E_B\;.
\end{equation}
As an example, we show in Fig.~\ref{fig6} 
the region of the classically allowed initial data for $\beta = \pi/3$, 
$\alpha = \pi/30$.
One observes that the function $N_{\mathrm{cr}}(E)$ oscillates 
between two 
linear envelopes, $E \cos^2(\beta+\alpha)$ and $E\cos^2(\beta-\alpha)$; 
the period of oscillations decreases as $E\to 0$. 
Moreover, the curve $N_{\mathrm{cr}}(E)$ has a number of minima at
the points $E = E_{n}^{\mathrm{cr}}$. This means that the energies 
${E = E_{n}^{\mathrm{cr}}}$ 
are optimal for reflection: in the vicinity of 
any point $E = E_{n}^{\mathrm{cr}}$, $N = N_{\mathrm{cr}}(E_{n}^{cr})$
reflections become exponentially suppressed
independently of whether 
the energy gets increased or decreased.
This feature is particularly pronounced in the case $\alpha+\beta=\pi/2$,
when the lower envelope coincides with the line $N=0$. Then, the classical 
reflections (i.e. reflections with the probability of order $1$) 
at $N=0$
are possible only in the vicinities of the points
\begin{equation}
\nonumber
E = \frac{1}{8\pi^2(n-1/2)^2}\;.
\end{equation}
This is the case we used in Introduction to illustrate
the effect. 

The 
minima $E = E_{n}^{\mathrm{cr}}$ exist at other values of the parameters
as well. For instance, let us find the positions of these minima in 
the case $\alpha\ll 1$. One differentiates 
Eq. \eqref{Ncr} with respect to energy and obtains,
\begin{equation}
\label{Emin}
E_{n}^{\mathrm{cr}} = E_{n}
\left\{ 1 - \frac{1}{\pi (n-1/2)} 
\mathrm{arcsin}\left(\frac{\mathrm{ctg}\, \beta}
{2 \pi \alpha (n-1/2)}\right) 
+ O(\alpha^2) \right\}\;,
\end{equation}
where 
\be
\label{En}
E_{n} = \frac{1}{8\pi^2(n-1/2)^2\sin^2\beta}
\ee
are the points where the curve $N = N_{\mathrm{cr}}(E)$ touches its
lower envelope. 
The argument of arcsine in Eq.~\eqref{Emin} should be smaller than one, 
so, the minima $E_{n}^{\mathrm{cr}}$
exist only at large enough $n$, 
\be
\label{n0}
n\geq n_0 
\equiv
\left[\frac{\mathrm{ctg}\, \beta }{2\pi\alpha}+\frac{1}{2}\right]+1\;,
\ee
where $[\cdot]$ stands for the integer part. 

Let us make several comments. First, note 
that $n_0 \sim O(1/\alpha)$, 
consequently, all the optimal points 
$E_{n}^{\mathrm{cr}}$ lie in the region of small energies
$E\sim 1/n_0^2 \sim O(\alpha^2)$. 
Second, as we pointed out before, the formula 
(\ref{Ncr}) for the function
$N_\mathrm{cr}(E)$ holds at $E>E_B$. Comparing the expressions
(\ref{En}), (\ref{n0}) and (\ref{EB}), one observes that 
$E_{n_0}>E_B$ if $\tg\beta>1$. So, there does exist a range of
energies where the non-monotonic behavior of the function
$N_\mathrm{cr}(E)$ can be inferred from the formula (\ref{Ncr}). In
fact, the conclusion about the existence of the local minima of
$N_\mathrm{cr}(E)$, as well as the expressions 
(\ref{Emin}), (\ref{En}), (\ref{n0})
determining
their positions, remain valid also at $E<E_B$. This follows from the
rigorous analysis of the boundary of the classically allowed region 
to which we
turn now. The reader who is more interested in the tunneling processes
may skip this part and proceed directly to 
subsection~\ref{sec:class-forb-refl}.   

Now, we do not appeal to the Ansatz
(\ref{Critical2}). Instead, we start with the general solution in the
intermediate region, 
\bseq
\label{eq:1}
\begin{align}
& x' = p'_0 (t - t_0)\;,\\
& y' = A'_0 \sin\left[(t-t_0)\cos\alpha +
    \varphi'\right]\;.
\end{align}
\eseq
It is convenient to parametrize it by the total
energy $E = p_0'^2/2 + \cos^2\alpha A_0'^2 /2$ and the ``inclination'' 
$\gamma$ defined by the relation
$$
p'_0/A'_0 = \mathrm{tg}\, \gamma \cos\alpha\;.
$$
Expressions (\ref{eq:1}) take the following form,
\bseq
\label{eq:2}
\begin{align}
&x' = \sqrt{2E}\; (t - t_0)\sin\gamma\;,\\
&y' = \sqrt{2E}\;\frac{\cos\gamma}{\cos\alpha} \sin
\left[(t - t_0) \cos\alpha + \varphi'\right]\;. 
\end{align}
\eseq
The constants $t_0$ and $\varphi'$ are fixed by demanding the
trajectory (\ref{eq:2}) to reflect classically from the second turn,
i.e. touch the line $\xi=0$ at $t = 0$,
$$
(x'-1)\cos\beta -y'\sin\beta\bigg|_{t=0} = 0\;,\qquad\qquad
\frac{d{y'}}{d{x'}}\bigg|_{t=0} =
\mathrm{ctg}\,\beta\;\;.
$$
These conditions imply,
\bseq
\label{eq:3}
\begin{align}
&t_0 = -\frac{1}{\sqrt{2E}\sin\gamma} +
  \frac{1}{\cos\alpha}
  \sqrt{\frac{\mathrm{tg}^2\,\beta}{\mathrm{tg}^2\,\gamma} - 1}\;,\\
&\varphi' = -\frac{\cos\alpha}{\sqrt{2E}\sin\gamma} +
  \sqrt{\frac{\mathrm{tg}^2\,\beta}{\mathrm{tg}^2\,\gamma} - 1} -
  \mathrm{arccos}\frac{\mathrm{tg}\, \gamma}{\mathrm{tg}\,\beta}\;.
\end{align}
\eseq
One sees that the classical reflections are possible only at $\gamma
\in [0;\,\beta]$; the boundary value $\gamma = \beta$ 
reproduces the solution (\ref{Critical2}).

In order to find $N_{\mathrm{cr}}(E)$, one should
minimize the value of the incoming oscillator excitation number with
respect to $\gamma$ at fixed $E$. At $t = t_0$, when the
particle crosses the first turn, 
\begin{equation}
\label{eq:4}
p_0 \equiv \dot{x}(t_0) = \sqrt{2E}
(\cos\alpha\sin\gamma - \sin\alpha \cos\gamma \cos \varphi')\;.
\end{equation}
Since $N = E - p_0^2/2$, one can maximize the value of the translatory
momentum $p_0$ instead of minimizing $N(\gamma)$.  
Formula (\ref{p0cr}) represents the value
$\gamma=\beta$ lying at the boundary of the accessible $\gamma$--domain;
this value should be compared to $p_0 (\gamma)$ taken at local maxima.

Let us consider the case $\alpha \ll 1$. At large enough energies,
$E\sim 1$, Eq. (\ref{eq:4}) is dominated by the first term,
which grows with $\gamma$, so that the maximum of
$p_0(\gamma)$ is indeed achieved at $\gamma=\beta$. At small energies,
however, the second term in 
Eq. (\ref{eq:4}) becomes essential because of the quickly oscillating
$\cos \varphi'$ multiplier: the frequency of $\cos
\varphi'$ oscillations grows as $E\to 0$, and at $E\sim\alpha^2$, 
in spite of the small
magnitude proportional to $\sin\alpha$, the second term 
produces the sequence of local maxima of the function $p_0(\gamma)$. 

\begin{figure}
\centerline{\includegraphics[width=0.7\textwidth]{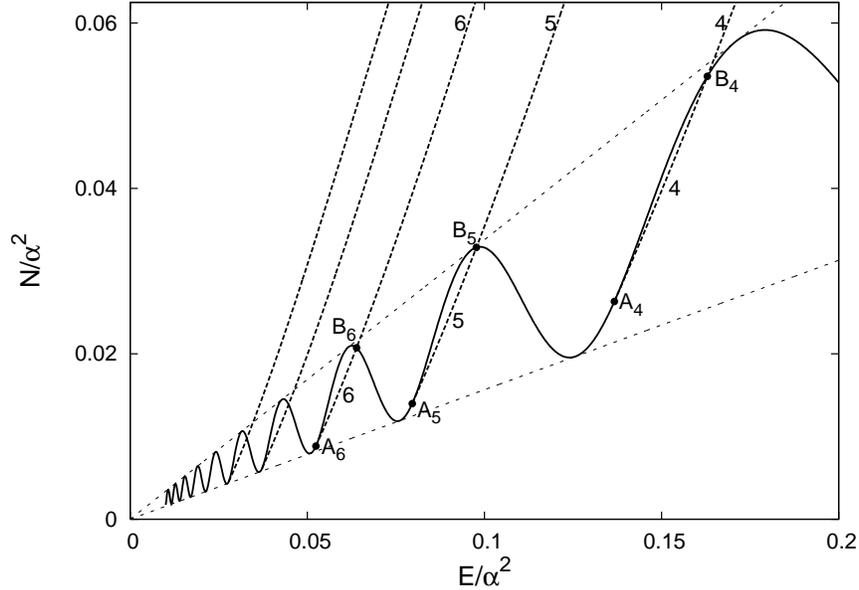}}
\caption{The graphs $N_n(E)$ corresponding to the local minima of the
  function $N(\gamma)$ (dashed lines) plotted together with the ``global''
  curve, Eq.~(\ref{Ncr}) (solid line); $\beta=\pi/3$,
  $\alpha=\pi/30$. The critical curve 
$N = N_{\mathrm{cr}}(E)$ is obtained by taking the minimum among
all the graphs.}
\label{Fig12}
\end{figure}

One expects the parameters of the trajectory at small $\alpha$
not to be very different from the ones at $\alpha=0$ (the latter case
was considered in Sec.~\ref{sec1turn}). So, we write,
$$
\gamma = \beta - \delta \gamma\;, 
$$
where $0<\delta \gamma \ll 1$. Expanding
the expressions (\ref{eq:3}), (\ref{eq:4}) and taking into account
that $E\sim \alpha^2$ one obtains,
\bseq
\label{eq:5}
\begin{align}
&\varphi' = -\frac{1}{\sqrt{2E}\sin\beta} (1 +
  \delta \gamma\,\mathrm{ctg}\,\beta)\;,\\  
&p_0 = \sqrt{2E} (\sin\beta - \delta \gamma \cos \beta -
  \alpha \cos\beta \cos \varphi')\;.
\end{align}
\eseq
Now, the local maxima of the initial translatory momentum can be
obtained explicitly by differentiating Eqs. (\ref{eq:5}) with respect
to $\delta \gamma$. One finds the sequence of them,
\begin{equation}
\label{eq:8}
 \delta \gamma_n = -\mathrm{tg}\, \beta + \sqrt{2E}\;
 \frac{\sin^2\beta}{\cos\beta} \left[2\pi n -\pi- \mathrm{arcsin}\left( 
\frac{\sqrt{2E} \sin^2\beta}{\alpha \cos\beta}\right)\right]\;.
\end{equation}
Only the maxima with $\delta\gamma_n>0$ should be taken into account.
The local maxima exist when 
\be
\label{EB}
E\leq E_B\equiv\frac{\alpha^2\cos^2\beta}{2\sin^4\beta}\;.
\ee
Substituting Eq.~(\ref{eq:8}) into the 
expressions (\ref{eq:5}), one evaluates the values of
$p_0$ at the local maxima,
\begin{equation}
\nonumber
\begin{split}
p_{0,n}(E) = &2\sqrt{2E}\sin\beta - 2E\sin^2\beta 
\left[2\pi n -\pi- \mathrm{arcsin}\left(
\frac{\sqrt{2E}\sin^2\beta}{\alpha \cos\beta}\right)\right]\\
 &+ \alpha
\sqrt{2E} \cos\beta \sqrt{1 - \frac{2E\sin^4\beta}{\alpha^2\cos^2\beta}}\;.
\end{split}
\end{equation}
The graphs $N_n(E) = E - p_{0,n}^2(E)/2$ are shown in 
Fig.~\ref{Fig12}  
for the case $\beta=\pi/3$, $\alpha=\pi/30$.
Each graph is plotted for
the  energy range $E>E_{A_n}$ restricted by the condition  $\delta
\gamma_n >0$. 
They are presented together with the curve given by the formula (\ref{Ncr}).
By definition, the critical solution corresponds to the
lowest of these graphs. Clearly, 
for
each ``local'' curve 
representing the $n$-th local minimum of $N(\gamma)$ 
there is a range of energies $E_{A_n} < E < E_{B_n}$ where it lies
{\it lower} than the ``global'' curve (\ref{Ncr}). This means
that the parameter $\gamma$ of the critical solution changes
discontinuously across the points $E = E_{B_n}$.
Correspondingly, the curve $N_\mathrm{cr}(E)$ has a break at these
points. 
On the other hand,  the function $N_{\mathrm{cr}}(E)$ is smooth at 
the points $A_n$ as   
the ``local'' graphs end up exactly at $\delta \gamma=0$, 
where the parameters of the $n$-th ``local'' solution coincide with
the ones of the ``global'' solution.

To summarize, we have observed that the
boundary of the classically allowed region is given by a collection of
many branches of 
classical solutions, each branch being relevant in its
own energy interval. We will see that a
similar branch structure is present in the complex trajectories 
describing 
over--barrier reflections in the classically forbidden region of $E$,
$N$.

\subsection{Classically forbidden reflections}
\label{sec:class-forb-refl}
In this subsection we demonstrate 
that the suppression exponent $F(E,\, N)$ 
viewed as a function of energy at fixed $N$  exhibits  
oscillations deep inside the classically forbidden region of initial
data. 
This result comes without surprise if one  takes into account the
non-monotonic behavior of the boundary $N_\mathrm{cr}(E)$ of the
classically allowed region. Indeed,
the
curve $N = N_{\mathrm{cr}}(E)$ coincides with the line $F(E,N) = 0$.
One has,
$$
\frac{dN_{\mathrm{cr}}}{dE} = -\frac{\partial_E F}{\partial_N F}
\Bigg|_{N = N_{\mathrm{cr}}(E)}\;,
$$
so that
$$
\frac{\partial F}{\partial E}(E_{n}^{\mathrm{cr}},\, 
N_{n}^{\mathrm{cr}}) = 0\;.
$$
We conclude that the points $E = E_{n}^{\mathrm{cr}}$ are  the 
local minima of the function $F(E)$ at fixed $N=N_n^\mathrm{cr}$. It
is natural to expect that such local minima of $F(E)$ exist at other
values of $N$ as well.  
To illustrate this fact explicitly,
we study the complex trajectories, solutions to 
Eqs.~\eqref{Tthetaeq}, \eqref{Tthetabc-1}, \eqref{Tthetabc+}. 

Following the tactics of the previous section, we find solutions in three 
separate regions: initial region $x'<0$, final region $\xi>0$, and the 
intermediate
region $x'>0$, $\xi<0$. These solutions, together with their first 
derivatives, should be glued  at $t = t_0$, when 
the complex trajectory crosses the line $x'=0$, 
and at $t = t_1$, when $\xi=0$. 
Besides, we are looking for the tunneling solution which ends up oscillating
along the line $AB$, see Fig.~\ref{fig5}. As discussed in
Sec.~\ref{sec1turn} this assumes existence of the second step of the
process: classical decay of the unstable orbit living at $\xi=0$; 
the latter decay is described by a real trajectory\footnote{One
  wonders why this trajectory does not 
  reflect from the turn $A'B'$ on its way back. This concern is
  removed by the observation that the trajectory produced in the decay
of the unstable orbit is not unique: in appendix we show that the
decay can occur at any point of the  segment $AC$ giving rise to a
whole bunch of potential decay trajectories. Most of these trajectories
pass through the turn $A'B'$ without
reflection.} going to $x\to-\infty$ at $t\to +\infty$.

The solution in the final region $\xi>0$ is (cf. Eqs. \eqref{theta1+}),
\bseq
\label{thetaeq+2}
\begin{align}
& \xi_+ (t) = 0\;,\\
& \eta_+(t) = \frac{\sqrt{2E}}{\cos\alpha\cos\beta} 
\sin(t\cos\alpha\cos\beta)\;,
\end{align}
\eseq
where we used the time translation 
invariance 
\eqref{TimeTranslations} to fix the final oscillator phase 
$\varphi_{\eta}=0$. In the intermediate region $x'>0$, $\xi<0$ one
writes, 
\bseq
\label{thetaeqint}
\begin{align}
\label{eq:7}
& x'(t) = p_0' t + x_0'\;,\\
\label{eq:9}
& y'(t) = a' \mathrm{e}^{-it\cos\alpha} + \bar{a}' 
\mathrm{e}^{it\cos\alpha}\;.
\end{align}
\eseq
Note that the final solution \eqref{thetaeq+2} does not contain
free 
parameters; thus, the matching of $x'$, $\dot{x}'$, $y'$, 
$\dot{y}'$ at $t = t_1$ enables one to express all the parameters in
Eqs. \eqref{thetaeqint} in terms of one complex variable $t_1$,
\bseq
\label{sew4}
\begin{align}
& p_0' = \sqrt{2E} \sin\beta \cos \phi_1 \;, \label{sew41}\\
& x_0' = 1 + \sqrt{2E} \frac{\mathrm{tg}\,\beta}{\cos\alpha} 
\left[ 
\sin\phi_1 - \phi_1 \cos\phi_1
\right]\;,\label{sew42}
\\
& a' = \frac{\sqrt{E/2}}{\cos\alpha}\mathrm{e}^{i\phi_1/\cos\beta}
\left[
\sin\phi_1 + i\cos\beta \cos\phi_1\right]\;,
\label{sew43}
\\
&\bar{a}' = \frac{\sqrt{E/2}}{\cos\alpha}\mathrm{e}^{-i\phi_1/
\cos\beta}
\left[
\sin\phi_1 - i\cos\beta \cos\phi_1\right]\;,
\label{sew44}
\end{align}
\eseq
where we introduced $\phi_1 = t_1\cos\alpha\cos\beta$. 

As the energy of the solution has been fixed  already, the only remaining 
initial condition involves initial oscillator excitation
number at $x' < 0$, see Eqs. \eqref{Tthetabc-1}. It is convenient 
to impose this condition at the matching point $t = t_0$. 
One recalls the definition of the matching time $t_0$,
$$
p_0' t_0 + x_0' = 0\;,
$$
which, after taking into account the 
expressions \eqref{sew41}, \eqref{sew42}, leads to the 
following equation,
\begin{equation}
\label{sewexact}
\frac{\cos\alpha}{\sqrt{2E}\sin\beta} + \frac{\sin\phi_1}{\cos\beta} - 
\cos\phi_1 \Delta  \phi = 0\;,
\end{equation}
where $\Delta\phi = \cos\alpha(t_1-t_0)$. At $t = t_0$
one has,
$$
\dot{x}(t_0) = p_0' \cos\alpha  - \dot{y}'(t_0)\sin\alpha  
= \sqrt{2(E-N)}\;,
$$
and thus
\begin{equation}
\label{Ns}
\frac{\sqrt{1-\nu}}{\sin\alpha} = \mathrm{ctg}\,\alpha \sin\beta\cos \phi_1
 - \sin \phi_1 \sin\Delta\phi - \cos\beta\cos\phi_1\cos\Delta\phi\;.
\end{equation}
As before,  $\nu = N/E$.

Two complex equations \eqref{sewexact}, \eqref{Ns} determine 
the matching times $t_0$, $t_1$, and, consequently, the complex 
trajectory. Although these equations cannot be solved explicitly, they
can be simplified in the case $\alpha \ll 1$, which we consider from
now on.
For concreteness, we study reflections 
at $N=0$.
It is important to keep in mind that 
 in the region of interest $E\sim E_{n}^{\mathrm{cr}}
\sim O(\alpha^2)$; thus, one should regard all the momenta
$p$ and oscillator amplitudes $a$, $\bar{a}$, as the quantities of 
order $O(\alpha)$. At the same time, for the distances along the 
waveguide one has  $x\sim O(1)$, so that the real parts of time intervals may
be parametrically large, $\mathrm{Re}\,t \sim x/p \sim O(1/\alpha)$. 

Further on, it will be convenient to work in terms of real variables, 
so, we represent $\phi_1$ and $\Delta\phi$ as 
$$
\phi_1 = \cos\alpha\cos\beta (\tau_1 + iT_1)\;,\qquad 
\Delta\phi = \cos\alpha(\tau + i\Delta T)\;.
$$
Note that $\tau$ and $\Delta T$ are 
the real and imaginary parts of the time interval $t_1-t_0$ which the
particle 
spends in the intermediate region.
Now, equation \eqref{sewexact} enables one to express
\begin{align} 
\label{sew5}
& \tau  = \frac{1}{\sqrt{2E} \sin\beta \mathrm{ch}(T_1\cos\beta)} + 
O(\alpha)\;, \\ 
&\label{sew6} \tau_1 = -\frac{1}{\tau\cos\beta}\left[
\frac{1}{\cos\beta} - \Delta T \mathrm{cth}(T_1\cos\beta)\right]
+ O(\alpha^3)\;.
\end{align}
Note that 
$\tau_1 \sim O(\alpha)$, $\tau \sim O(1/\alpha)$.
Then, the real part of Eq. \eqref{Ns} implies that
\be
\label{T1}
\mathrm{ch}(T_1\cos\beta) = \frac{1}{\sin\beta}
\left[1+\alpha \, \mathrm{ctg}\beta \cos\tau \mathrm{e}^{\Delta T}
\right] + O(\alpha^2)\;.
\ee
While deriving this formula we imposed $T_1<0$ which follows from the
requirement that in the limit $\alpha\to 0$ equation~(\ref{sew1_1}) should
be recovered; besides, we 
assumed $\e^{\Delta T}\sim O(1)$.  
Substituting Eq.~(\ref{T1}) into Eq.~\eqref{sew5} and
the imaginary part of 
Eq.~\eqref{Ns}, we obtain the final set of equations,
\bseq
\label{1-2}
\begin{align}
\label{1} & 1-\tau \sqrt{2E} = \alpha \, \mathrm{ctg}\beta \cos\tau 
\mathrm{e}^{\Delta T} + O(\alpha^2)\;,\\
\label{2} & (1+\Delta T) \mathrm{e}^{-\Delta T} = 
\alpha \, \mathrm{ctg}{\beta}\tau\sin\tau + O(\alpha)\;.
\end{align}
\eseq
These two nonlinear equations, still, cannot be solved explicitly. 
Nevertheless, one 
can get a pretty accurate idea about the structure of their 
solutions.

Before proceeding to the analysis of the above equations, let us 
derive a convenient expression for the suppression exponent 
$F_0(E)\equiv F(E,N=0)$.
Note that on general grounds 
one expects to obtain an expression of
the form,
\be
\nonumber
F_0(E)=E(f_\beta(0)+O(\alpha))\;,
\ee 
where $f_\beta(0)$ is given by Eq. (\ref{f0}). We are interested in the
$O(\alpha)$ 
correction in this expression, so, one must be 
careful to keep track of the subleading terms during the derivation. 

Making use of the equations of motion, one obtains for the incomplete
action (\ref{Sbyparts}) of the
system,
\begin{equation}
\nonumber
2\Im\, \tilde{S}  = \Im\; p_0' 
 = \sqrt{2E}\sin\beta\;\Im(\cos \phi_1)\;.
\end{equation}
Substitution of Eqs.~(\ref{sew5}), (\ref{sew6}),
(\ref{T1}) into this formula yields
\be
\nonumber
2\Im\tilde S=2E\left\{-1-\Delta T-\alpha\ctg\beta\cos\tau\e^{\Delta T}
\left(1+\frac{1}{\cos^2\beta}+2\Delta T\right)
+O(\alpha^2)\right\}\;.
\ee
For the parameter $T$ one has (see Eqs. (\ref{Tthetabc-2})),
\be
\label{T_1}
T=-\frac{2\Im x_0}{p_0}=-\frac{2\Im(x(t_0)-p_0t_0)}{p_0}=
2(T_1-\Delta T)+\sqrt{\frac{2}{E}}\sin\alpha\Im y'(t_0)\;,
\ee
where in the last equality we used Eqs.~(\ref{xsysxieta}) and
$x'(t_0)=0$. The quantity  
$\Im y'(t_0)$ is evaluated by using Eqs. (\ref{eq:9}), (\ref{sew4})
and (\ref{T1}); one finds,
\be
\nonumber
\Im y'(t_0)=-\sqrt{2E}\big(\ctg\beta\cos\tau\e^{\Delta T}+O(\alpha)\big)\;.
\ee
Substituting everything into the formula (\ref{F}), we obtain,
\be
\label{Falpha}
F_0(E)=E\Big(f_\beta(0)-4\alpha\ctg\beta\cos\tau\,\Delta T\e^{\Delta T}
+O(\alpha^2)\Big)\;.
\ee 
This expression implies that determination of the $O(\alpha)$
correction to the suppression 
exponent involves finding $\tau$, $\Delta T$ with
$O(1)$--accuracy. 
This is precisely the level of
accuracy of Eqs.~(\ref{1-2}). Below we will also need the
following formulae, which can be easily obtained by using 
$T=-\frac{\d F}{\d E}$ and Eq.~(\ref{T_1}),
\begin{align}
\label{dFdE}
&\frac{dF_0}{dE}=f_{\beta}(0)+2(\Delta T+1) +O(\alpha)\;,\\
\label{dFEdE}
&\frac{d}{dE}\,\left(\frac{F_0}{E}\right)=\frac{2(\Delta T+1 +O(\alpha))}{E}\;.
\end{align}
Note that, though  the suppression exponent differs from that in the one--turn
case only by $O(\alpha)$ correction, its derivative gets modified in the zeroth
order in $\alpha$. 

\begin{figure}
\centerline{\includegraphics[angle=-90,width=0.75\textwidth]{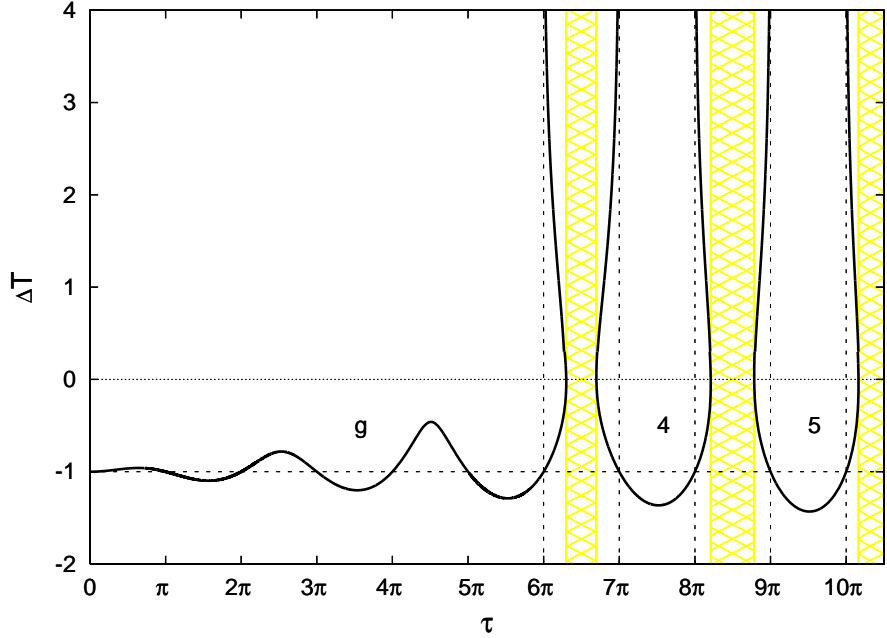}}
\caption{Curves representing solutions to Eq.~(\ref{2});
  $\beta=\pi/3$, $\alpha=\pi/30$.}
\label{fig13}
\end{figure}

Now, we are ready to analyze  Eqs. \eqref{1-2}. One
begins by solving  Eq. (\ref{2}) graphically, see
Fig. \ref{fig13}. The important property 
of this equation is as follows. 
One notices that the l.h.s. of Eq. (\ref{2}) 
is always smaller than $1$, the maximum being achieved at 
$\Delta T = 0$. Therefore, the solutions to this equation are confined to
 the bands
$$
\tau \sin\tau < \frac{\mathrm{tg}\beta}{\alpha}\;.
$$
This corresponds to
\begin{equation}
\label{bands}
\tau \in [0;\,2\pi (n_1-1)+\delta \tau_{n_1}] 
\qquad\mathrm{or} \qquad
\tau \in [2\pi n-\pi - \delta \tau_n ;\,2\pi n+\delta \tau_n]\;,
~~~n\geq n_1
\end{equation}
where 
\begin{align}
\notag
&\delta \tau_n = \mathrm{arcsin}\left(\frac{\mathrm{tg}\beta}{2\pi
 \alpha(n-1/2)} 
\right) + O(\alpha)\;,\\
&n_1=\left[\frac{\mathrm{tg}\beta}{2\pi\alpha} +
  \frac12\right]
 + 1\;,
\label{n1}
\end{align}
with $[\cdot]$ in the last formula 
standing for the integer part. 
The forbidden bands, where $\tau \sin\tau > \mathrm{tg}\beta/\alpha$,
are marked in Fig. \ref{fig13}  by yellow shading. The property
(\ref{bands}) introduces a 
topological classification of the solutions $\tau$, $\Delta T$ to 
Eqs.~(\ref{1-2}).
Namely, these solutions fall into a set of continuous branches: the
``local'' branches $\tau_n(E)$, $\Delta T_n(E)$ living inside the
strips 
$\tau \in [2\pi n-\pi - \delta \tau_n ;\,2\pi n+\delta\tau_n]$, 
$n\geq n_1$, and the ``global'' branch $\tau_g(E)$, $\Delta T_g(E)$
inhabiting the very first band 
$\tau \in [0;\,2\pi (n_1-1)+\delta \tau_{n_1}]$.
As follows from the definition of $\tau$,
the topological number $n$ counts the 
number of $y'$--oscillations during the evolution in the intermediate 
region.

\begin{figure}
\centerline{\includegraphics[angle=-90,width=0.8\textwidth]{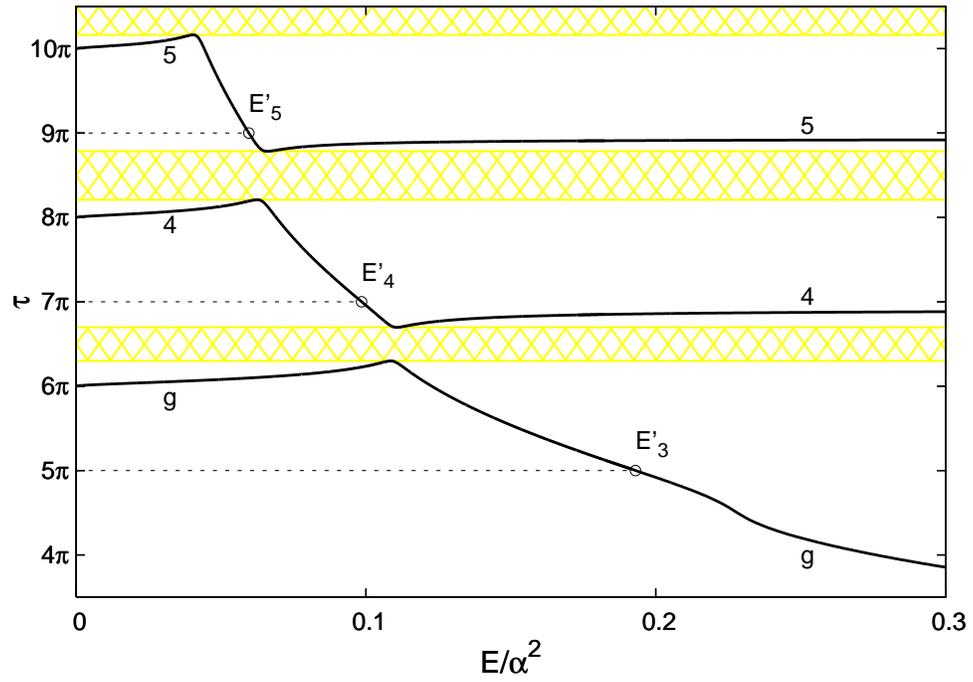}}
~\\
~\\
\centerline{\includegraphics[angle=-90,width=0.8\textwidth]{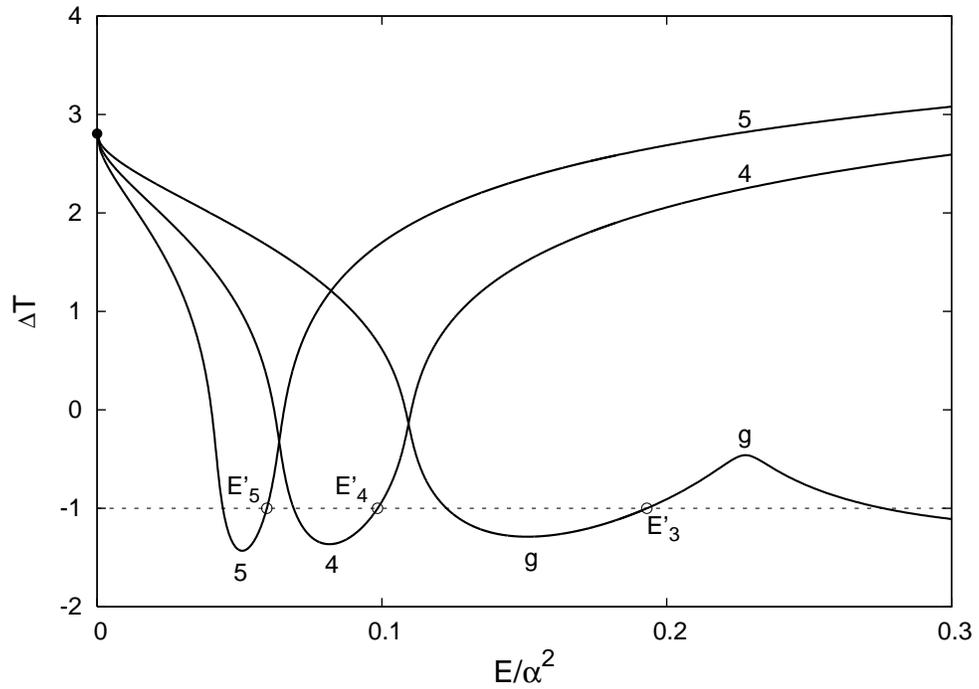}}
\caption{Several first branches of solutions to Eqs.~(\ref{1-2}): 
``global'' branch (``g'') and two ``local''
  branches (``4'', ``5''); $\beta=\pi/3$, $\alpha=\pi/30$.}
\label{fig7}
\end{figure}

Let us consider the ``global'' branch. From Eqs.~(\ref{1-2})
one has,
\begin{align}
\notag
&\tau_g\to 2\pi(n_1-1)+O(\alpha\ln\alpha)\;,&
&\Delta T_g\to\ln(\tg\beta/\alpha)\;,&
&E\to 0\;,\\
\notag
&\tau_g\to 0\;,&
&\Delta T_g\to -1\;,&
&E\to +\infty\;.
\end{align}
By inspection of Fig.~\ref{fig13} one can work out the
qualitative behavior of the functions $\tau_g(E)$, $\Delta
T_g(E)$. Alternatively, these functions can be found numerically. 
They are plotted in Fig.~\ref{fig7} for the case
$\beta=\pi/3$, $\alpha=\pi/30$
(the curves marked with ``g''). 
One observes that at high enough energies the function $\Delta
T_g(E)$ exhibits oscillations around the line $\Delta T=-1$. According
to the formula (\ref{dFEdE}) this means that the function $F_0(E)/E$
is non-monotonic, it attains local minima at the points
\begin{equation}
\label{Eoptimal}
E'_{n} = \frac{1}{8\pi^2 (n-1/2)^2}\left[ 1 + 
2\alpha\mathrm{e}^{-1}\mathrm{ctg}\beta + O(\alpha^2)\right]\;.
\end{equation}
Moreover, if 
\be
\label{n0prime}
n\geq n_0'\equiv\left[\frac{\tg\beta}{4\pi\alpha}\,
f_\beta(0)\exp\left(1+\frac{f_\beta(0)}{2}\right)
+\frac{1}{2}\right]+1 
\ee
there exist $E_n^o=E_n'(1+O(\alpha))$, such that 
$\Delta T(E_n^o)=-1-f_\beta(0)/2$. Then, according to Eq. (\ref{dFdE})
the points $E_n^o$ are the ``optimal'' energies corresponding
to the local minima of the suppression
exponent $F_0(E)$. 

\begin{figure}
\centerline{\includegraphics[angle=-90,width=0.7\textwidth]{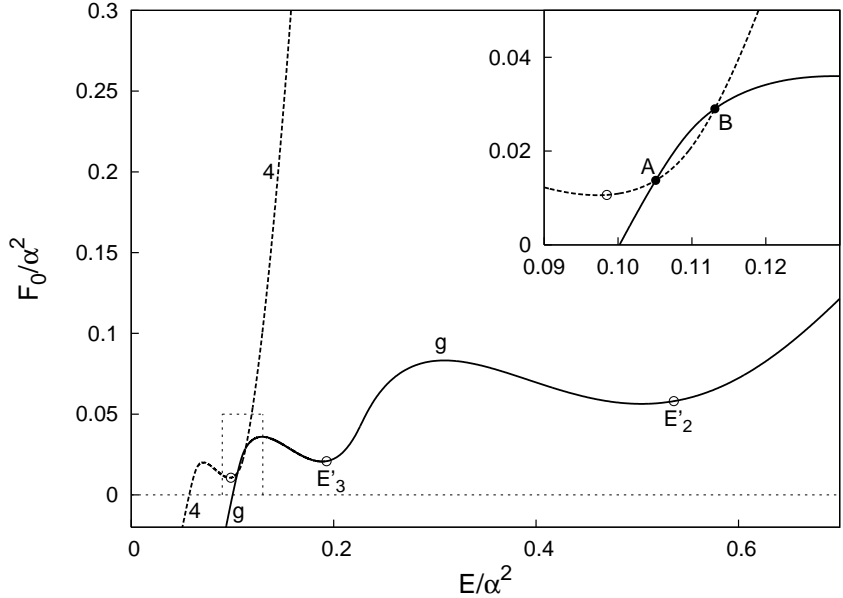}}
\caption{\label{fig10} The suppression exponent $F_0(E)$ 
for the ``global'' and first ``local'' ($n=4$) branches;
$\beta=\pi/3$, $\alpha=\pi/30$. 
The vicinity of intersection of the graphs is 
enlarged in the upper right corner. }
\end{figure}

At low energies the function $\Delta T_g(E)$
ceases to oscillate and becomes large and positive. According to
Eq.~(\ref{dFdE}) this means that the suppression exponent $F_{0,g}(E)$
of the ``global'' solution becomes negative at low 
energies\footnote{It is worth mentioning that Eqs. (\ref{1-2}) and the
  expression (\ref{Falpha}) for the suppression exponent become
  inapplicable at large $\Delta T$: the assumption $\e^{\Delta T}\sim O(1)$
  which was used in the derivation of these equations
gets violated. Nevertheless, by analyzing the full equations (\ref{sewexact}),
(\ref{Ns}) one can show that $dF_{0,g}/dE= -T_g$ is large and positive   at
$E\to 0$. This is sufficient for concluding that 
$F_{0,g}(E)$ is negative in the low--energy domain.}, 
see Fig.~\ref{fig10}. This is a clear signal that the ``global''
solution becomes unphysical at these energies and its contribution to
the reflection probability should be discarded: negative suppression
exponent contradicts the unitarity requirement\footnote{Another
  indication that the 
  ``global'' solution is unphysical at small $E$ is that the function
  $\tau_g(E)$ is
  bounded from above.
Indeed, $\tau$ is the time interval the
  particle spends in the intermediate part of the waveguide,  one
  expects it to  tend to infinity as $E\to 0$ for a physically
  relevant solution.}, ${\cal P}<1$.
One is forced to conclude that at low energies reflection is described
by the ``local'' solutions. Let us study them in detail.

For the $n$-th branch one obtains,
\begin{align}
\notag
&\tau_n\to 2\pi n+O(\alpha\ln\alpha)\;,&
&\Delta T_n\to\ln(\tg\beta/\alpha)\;,&
&E\to 0\;,\\
\notag
&\tau_n\to 2\pi n -\pi\;,&
&\Delta T_g\to +\infty\;,&
&E\to +\infty\;.
\end{align}
From  Fig.~\ref{fig13} one learns 
that the $n$-th solution
passes through the points
\begin{equation}
\label{eq:10}
\Delta T_n = -1\;, \qquad \tau = 2\pi n\qquad \mathrm{or}
\qquad\tau = 2\pi n - \pi\;.
\end{equation}
Thus, each curve $\Delta T_n(E)$ has one sharp dip, its minimum is 
smaller than $-1$, see Fig.~\ref{fig7}. 
As in the case with the ``global'' branch, the points (\ref{eq:10})
represent the 
extrema of the functions $F_{0,n}(E)/E$; the positions of the
local minima are again given by Eq.~(\ref{Eoptimal}).

Making use of Eq.~(\ref{Falpha}), we find that the suppressions
$F_{0,n}(E)$ of the ``local'' branches are large and positive at high
energies. Hence, these solutions give subdominant contributions to the
reflection probability at such $E$ as  
compared to the ``global'' solution. As energy decreases, $F_{0,n}(E)$
also decreases, then makes one oscillation and drops to negative values at
small $E$. The latter property means that each ``local'' branch
becomes unphysical at small enough energies. The suppression exponent of
the first ``local'' branch (corresponding to $n=4$ in the case
$\beta=\pi/3$, $\alpha=\pi/30$) is presented in Fig.~\ref{fig10}.

An alert reader may have already guessed that we have met here 
the typical Stokes phenomenon~\cite{Elyutin2001}. 
In fact, the Stokes phenomenon is specific to the situations where some
integral (e.g., the path integral (\ref{AA}) in our case) is
evaluated by the saddle--point method. Essentially, it
means the following: as one gradually changes the parameters of the
integral in question, 
a given saddle point may become {\it non--contributing}
after the values of these parameters cross a certain curve drawn in the
parameter space, the Stokes line. Since the
result of the computation should be continuous, this phenomenon 
occurs only for subdominant saddle points (saddle--point trajectories in
our case). 
Unfortunately, apart from several heuristic
conjectures~\cite{Elyutin2001,Stokes}, sometimes rather 
suggestive~\cite{Stokes-Shudo}, there is presently no 
general method of dealing with 
the Stokes phenomenon in
the semiclassical calculations.
However, in the situation encountered
above it suffices to use the simplest
logic lying at the heart of all other 
approaches\footnote{The simplification in the present case is
  related to the fact that we concentrate on the dominant
  semiclassical contribution, leaving aside the subdominant ones.}.
 
\begin{figure}[htb]
\centerline{\includegraphics[angle=-90,width=0.8\textwidth]{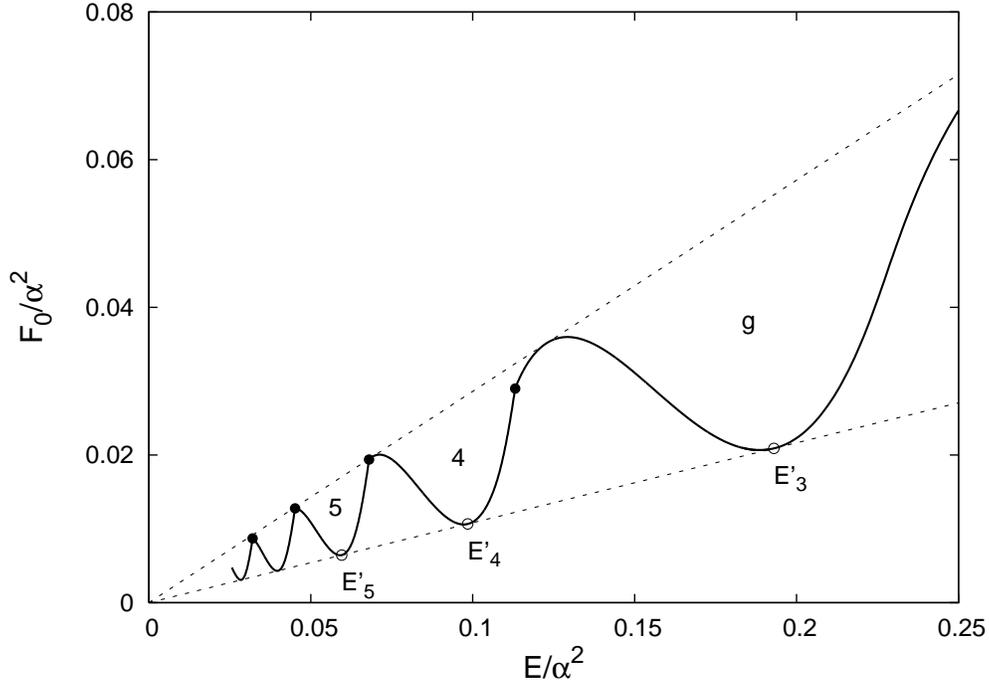}}
\caption{The final result for the suppression exponent 
$F_0(E)$
in the region of small energies; $\beta=\pi/3$, $\alpha=\pi/30$. 
The points where different branches 
merge are shown with thick black dots.}
\label{fig9} 
\end{figure}

When gathering the final result for the suppression exponent, we 
follow two guidelines. First, it is clear that, as energy
decreases, each branch becomes unphysical {\it before}
 $F_{0,n}(E)$ crosses zero. 
On the other hand, at high energies one 
should pick up the branch corresponding
to the smallest value of the suppression exponent. Looking 
at Fig.~\ref{fig10},
one notes that the curves $F_{0,g}(E)$, $F_{0,4}(E)$ have two 
intersections, $A$ and $B$. At $E > E_B$ one chooses the ``global'' 
branch. In the region $E_A < E < E_B$ we switch to the first ``local'' 
branch, because in this region $F_{0,4}(E) < F_{0,g}(E)$. Naively, at 
$E = E_A$ one should jump back to the ``global'' branch; however, 
in order to preserve unitarity at small energies, we suppose that somewhere
in between the points $B$ and $A$ the ``global'' branch becomes 
non--contributing, so that one should stay at the ``local'' 
branch at $E < E_A$. 
Similarly, the adjacent ``local'' branches have two intersections; 
as the energy decreases, we switch from $n$-th branch to $n+1$-th at
the first intersection, and stay there until the 
intersection with the $n+2$-th
branch. Overall, one obtains the graph for the suppression exponent 
plotted in Fig.~\ref{fig9}. 
The suppression 
exponent oscillates between two linear envelopes, 
$F=E(f_\beta(0)\pm 4\e^{-1}\alpha\ctg\beta)$; oscillations 
pile up in  the region of low energies. The 
reflection process is optimal in the vicinities of the minima of the
function 
$F_0(E)$. 

\section{Discussion}
\label{sec:discussion}
By considering a class of two--dimensional waveguide models, we have
demonstrated explicitly that the probability of over--barrier
reflection can be
non--monotonic function of energy.
The origin of the effect lies in the classical dynamics: the
parameters of the complex trajectory describing over--barrier
reflection change quasi-periodically as the energy gets
decreased. This results in the oscillatory behavior of the
suppression exponent. Reflection occurs with exponentially larger
probability in the vicinities of ``optimal'' energies (local minima
of the suppression exponent) while being highly suppressed in
between. 

Our results are obtained for a fairly 
specific class of waveguides, namely, the ones
with very sharp turns. However, the qualitative features observed in this 
paper  
should be valid for quite general waveguide models:
a classical particle with high energy feels any large--scale
turn of the waveguide as a sharp one\footnote{More precisely,
 one should compare the width $b$ of the turn to the quantity
 $\frac{2\pi}{\omega}\frac{p_0}{m}$, where $p_0$ is the translatory
 momentum of the particle and $\omega$ stands for the frequency of
 transverse oscillations; if $b\ll \frac{2\pi p_0}{\omega m}$, one
 is in the class of models with sharp turns.}; if two turns are separated by 
a long  interval  of free motion, one arrives to the model (\ref{a2}). 
We remark  that the  phenomenon of optimal tunneling 
has been observed also in numerical 
investigation of a  smooth waveguide, see
Ref.~\cite{Levkov:2007ce}.  

The branch structure of solutions observed in the 
region of small energies is  interesting 
from the mathematical point of view. We have shown that there 
exists an infinite sequence of complex trajectories marked by 
the topological number $n$. Each branch produces physically consistent
result for the suppression exponent in some energy 
interval; 
outside of this interval the $n$-th
branch would correspond either to highly suppressed transitions (high
energies) or to violation of unitarity (low energies).   
We collected the final graph for the
suppression  
exponent basing on the empirical considerations, which hardly may be
acknowledged 
as satisfactory. Our study clearly shows that the method of complex 
trajectories should be equipped with a convenient rule to pick up the
physical trajectory 
among the discrete set of solutions to the boundary value problem 
\eqref{Tthetaeq}, \eqref{Tthetabc-1}, \eqref{Tthetabc+} (in other
words, the method
to deal with the Stokes phenomenon). Presently, 
such a rule is absent.

We note that the described physical 
phenomenon of optimal tunneling is present independently of the way 
the branches of solutions are glued together. The result at relatively 
high energies is given by the ``global'' branch, which displays
a large number of local minima if
$n_0'>n_1$, see Eqs. (\ref{n0prime}),
(\ref{n1}). This is the case for the illustrative example considered
throughout this paper, see Fig.~\ref{fig10}.  

As a final remark, we point out some open issues. We have calculated the 
suppression exponent of reflection using the sharp--turn
approximation. It would be instructive to extend our analysis
by finding corrections due to the finite turn widths. The motivation
is twofold. First, the analysis performed in appendix implies
existence of a rich variety of distinctive semiclassical solutions
contributing almost equally into the reflection probability.
This feature might be a manifestation of
chaos~\cite{chaotic} which is present in our system but hidden by
the sharp--turn approximation. [Note that chaos is inherent in a very similar
  waveguide model with smooth potential, see
  Ref.~\cite{Levkov:2007ce}.] Clearly, the structure 
of solutions in the vicinities of the turns is worth further
investigation. 

Second, it was proposed recently in
Refs.~\cite{Bezrukov:2003,Mechanism-Takahashi} that
the process of dynamical tunneling in quantum systems with multiple
degrees of freedom (including field theoretical models, see
Refs.~\cite{Rubakov:1992ec}) can proceed differently from the
ordinary case of one--dimensional tunneling. Namely, classically
unstable state can be created during the process; this
state decays subsequently into the final asymptotic region. 
The analysis performed in the present paper naturally conforms with
this tunneling mechanism: all our complex trajectories are
matched with the unstable orbit living at the turn.
Still, the sharp--turn approximation does not allow to distinguish
between the truly unstable trajectories staying at the turn forever and
those which reflect from the turn in a finite time.
To decide whether the tunneling mechanism of 
Refs.~\cite{Bezrukov:2003,Mechanism-Takahashi}
is indeed realized in our model one needs to go beyond the sharp--turn
approximation. Then, 
the candidate for the ``mediator'' unstable state  is 
the ``excited sphaleron'', the solution considered in the appendix. 
Presumably, in our model one can answer analytically to the
question of whether or not the ``excited sphaleron'' acts
as an intermediate state of the tunneling process.
 This study is quite beyond the scope of the present
paper and we leave it for future investigations.

\paragraph*{Acknowledgments.}
We are indebted to F.L.~Bezrukov and V.A.~Rubakov for the encouraging
interest and helpful suggestions. This work is supported in part by
the Russian Foundation for Basic Research, grant 05-02-17363-a;
Grants of the President of Russian Federation NS-7293.2006.2
(government contract 02.445.11.7370), MK-2563.2006.2 (D.L.),
MK-2205.2005.2 (S.S.); Grants of the Russian Science Support Foundation
(D.L. and S.S.); the personal fellowship of the ``Dynasty''
foundation (awarded by the Scientific board of ICFPM) (A.P.) and
INTAS grant YS 03-55-2362 (D.L.). D.L. is grateful to Universite
Libre de Bruxelles and EPFL (Lausanne) for hospitality during his
visits.

\appendix
\section{Classical motion near the turn}
\label{appA}
In this appendix we analyze the  motion of the particle near the sharp 
turn of the waveguide (\ref{a1}) at nonzero smoothening of the turn,
see, e.g., Eq.~\eqref{thetab}. 
We suppose that in the small vicinity of the turn the 
function $w(\xi, \, \eta)$ can be represented in the form
\begin{equation}
\label{repr}
w(\xi,\, \eta) = \cos\beta\left(\eta  - bv(\xi/b)\right)\;,
\end{equation}
where $v(\psi)$ does not depend explicitly on $b$.
Moreover, we consider the case when $v(\psi)$ has a 
maximum\footnote{For the smoothening \eqref{thetab}, the properties 
\eqref{repr}, \eqref{amin} hold with $v(\psi) = \frac{\psi \mathrm{tg}\beta}
{1+\mathrm{e}^{\psi}}$, $\psi_0 \approx 1.28$.},
\begin{equation}
\label{amin}
v'(\psi_0) = 0\;.
\end{equation}

Due to the property \eqref{amin} one immediately obtains the exact 
periodic solution to the equations of motion \eqref{Tthetaeq}, which
we call ``excited sphaleron''~\cite{Bezrukov:2003},
\be
\label{Sphaleron}
\xi_{\mathrm{sp}} = b\psi_0\;,\qquad
\eta_{\mathrm{sp}} = A_{\eta}\sin( t\cos\beta + \varphi_{\eta}) + 
bv(\psi_0)\;.
\ee
We are going to show that this solution is unstable: a small
perturbation above it grows with time and the particle flies away to
either end of the waveguide. In particular, there are solutions
that  describe the decay of the sphaleron to  $\xi\to -\infty$ both at
$t\to\pm \infty$. Clearly, 
such solutions correspond to reflections from the turn.

In the vicinity of the sphaleron the trajectory of the particle can be
represented in the form, 
\be
\label{vic}
\xi = b\psi(t)\;,\qquad
\eta = \eta_{\mathrm{sp}}(t) + b\rho(t)\;,
\ee
where $\psi,\,\rho \sim O(1)$. Writing down the classical equations of motion 
\eqref{Tthetaeq} in the leading order in $b$, one obtains,
\begin{align}
\label{psi}
&\frac{d^2\psi}{ds^2} = \frac{4}{b} A_\eta \sin(2s) v'(\psi)\;, \\
\label{rho}
& \frac{d^2\rho}{ds^2} + 4 \rho = {4}[v(\psi)-v(\psi_0)]\;,
\end{align} 
where $s = (t\cos\beta  +\varphi_\eta)/2$. It is worth noting that 
the right hand side
of Eqs. \eqref{psi}, \eqref{rho} are of different order in $b$. We
will see that due to this difference $\rho=0$ in the leading order in $b$.

Let us first consider the linear perturbations above the excited sphaleron, 
$$
\psi = \psi_0 + \delta \psi\;,\qquad \delta\psi\ll 1\;.
$$
Equation \eqref{psi} can be linearized with respect to $\delta \psi$
leading to the Mathieu equation
\begin{equation}
\nonumber
\frac{d^2}{ds^2}\delta\psi + 2 q \sin(2s)\delta\psi = 0\;,
\end{equation}
with canonical parameter $q = -2v_0'' A_\eta/b > 0$. 
As $q \sim O(1/b)\gg 1$, one can apply the WKB formula,
\begin{equation}
\label{deltapsi}
\delta\psi = \frac{A\cos W}{\sqrt{dW/ds}}\;,
\end{equation}
where $|A|\ll 1$, and 
\begin{equation}
\nonumber
W = \sqrt{2q} \int_{\pi/4}^{s} ds'\, \sqrt{\sin(2s')}\;.
\end{equation}
Note that we have chosen the solution symmetric with respect to time 
reflections,
\begin{equation}
\label{AssSymm}
\delta\psi(\pi/2-s) = \delta\psi(s)\;.
\end{equation}
At $s \in [0;\,\pi/2]$ the exponent $W$ is real 
and the particle gets stuck at $\psi\approx \psi_0$, oscillating
around this point with high frequency $dW/ds \sim O(b^{-1/2})$.
At $s<0$ the solution~(\ref{deltapsi}) grows
exponentially, meaning that the particle flies away from the excited
sphaleron, 
\be
\nonumber
\delta\psi (s<0) = \frac{A\cos(W(0)-\pi/4)}{\sqrt{|dW/ds|}}
\mathrm{e}^{|W(s)-W(0)|}\;.
\ee
In what follows, we choose
$A\cos(W(0)-\pi/4) < 0$,
so that $\delta\psi < 0$ at  $s < 0$. Let us denote by $s_1<0$ the
point where $\delta\psi$ becomes formally equal to $-1$,
$$
\frac{A\cos(W(0)-\pi/4)}{\sqrt{|dW/ds|}}
\mathrm{e}^{|W(s_1)-W(0)|}=-1\;. 
$$
In what follows we suppose that $s_1 \sim O(1)$, hence, $A$ is
exponentially small.  
Then, in the vicinity of this point, $|s-s_1|\ll 1$, one has,
\be
\label{psis1}
\delta\psi=-\exp{\left(\sqrt{-2q\sin(2s_1)}(s_1-s)\right)}
=-\exp{\left(\sqrt{4v''_0A_\eta\sin(2s_1)}
\;\frac{(s_1-s)}{\sqrt b}\right)}\;.
\ee
We notice that $\delta\psi$ evolves from exponentially small values to
$\delta\psi\sim O(1)$ during the characteristic time 
$|s-s_1|\sim O(\sqrt b)$. 

\begin{figure}
\centerline{\includegraphics[angle=-90,width=0.6\textwidth]{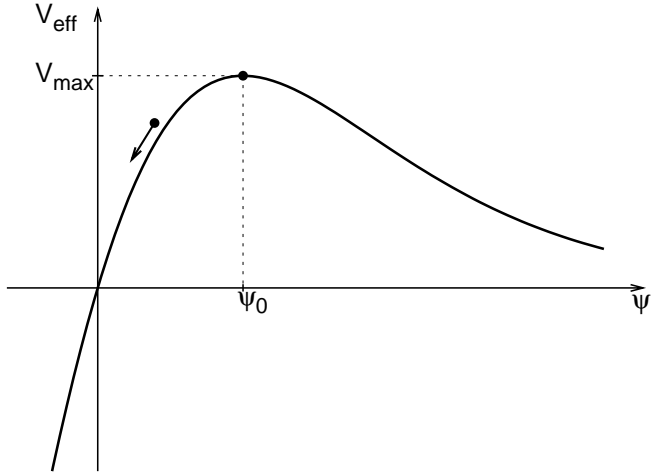}}
\caption{The effective potential for Eq. \eqref{Nonsolved}. }
\label{fig3}
\end{figure}

When $\delta\psi\sim O(1)$ the linear approximation breaks
down and one has to solve the nonlinear equation (\ref{psi}). Using
$s=s_1+O(\sqrt b)$ one writes
\be
\label{Nonsolved}
\frac{d^2\psi}{ds^2} = \frac{4}{b} A_\eta \sin(2s_1) v'(\psi)\;.
\ee
This equation permits to draw a useful analogy with one--dimensional
particle moving in the effective potential 
$V_{eff}(\psi)=-4b^{-1}A_\eta\sin(2s_1)v(\psi)$ (see Fig.~\ref{fig3}).
This auxiliary particle starts in the 
region near the maximum of the potential at $(s-s_1)/\sqrt b\to +\infty$
with energy $E\approx V_{max}$ and rolls down  toward 
$\psi \to -\infty$ at $(s-s_1)/\sqrt b\to -\infty$. 
In this limit $v(\psi)\to\psi\tg\beta$ and the solution takes the form
\be
\nonumber
\psi=C_1+C_2(s-s_1)+2b^{-1}A_\eta\sin(2s_1)\tg\beta\; (s-s_1)^2\;.
\ee
Note that the coefficients $C_1$, $C_2$ here are not independent: they
are determined by the
parameter $s_1$ through matching of the solution with Eq.
(\ref{psis1}) at $(s-s_1)/\sqrt b\to+\infty$. We do not need
their explicit form, however. 

Let us argue that the function $\rho$ remains small during the
whole evolution of the particle in the vicinity of the
sphaleron. Indeed, in the linear regime one has $\delta\psi\ll 1$ and
the r.h.s. of Eq.~(\ref{rho}) is small. So, $\rho$ does not get excited. On
the other hand, the nonlinear evolution of $\psi$ proceeds in a short
time interval $\Delta s=O(\sqrt b)$; so, again, $\rho$ is suppressed
by some power of $b$.

The trajectory (\ref{vic}) found in the vicinity of the sphaleron
should be matched at 
$$
1\gg |s-s_1|\gg\sqrt b
$$
with the free solution in the asymptotic region
$\xi<0$, see Eqs.~(\ref{ClassAllTraj}). 
It is straightforward to check that matching can be
performed up to the second order in $(t-t_1)$, which is consistent with
our approximations. In this way one determines the free asymptotic
solution which, up to corrections of order $O(b)$, coincides with the
sinusoid coming from $\xi\to -\infty$ at $t\to -\infty$ and touching
the line $\xi=0$ at $t=t_1$.

Now we recall that, by construction, the obtained solution is
symmetric with respect to time reflections,
$$
\xi(s)=\xi(\pi/2-s)\;,\qquad \eta(s)=\eta(\pi/2-s)\;.
$$
This means that it satisfies $\xi\to -\infty$ at $t\to \pm\infty$.
This solution describes reflection of the particle from the
turn. 
 
The reasoning
 presented in this appendix puts considerations of the main body of
 this paper on the firm ground: we have found the ``smoothened''
 solutions which reflect classically from the
 turn, and in the limit $b\to 0$ coincide with the free solutions 
of Sec.~\ref{sec1turn} touching the line $\xi=0$.

It is worth mentioning that, apart from the reflected solution we
have  found, in the vicinity of any trajectory
touching the line $\xi=0$ there exists 
a rich variety of qualitatively different
motions.
First of all, one may 
successfully search for solutions which are odd with respect to time 
reflections  (Eq. \eqref{AssSymm} with minus sign). Such solutions,
though close to the reflected ones at $t<0$,  
describe transmissions of the particle through 
the sharp turn into the asymptotic region $\xi \to +\infty$. 
Relaxing the time reflection symmetry,
one can find solutions leaving the vicinity of the turn at 
any point $\eta<0$, which is different, in general, from 
the starting point $\eta =
\eta(s_1)$. 
Yet another types of solutions are obtained in the
case when  the amplitude $A$ of  $\delta \psi$--oscillations at $s \in
[0;\; \pi/2]$ is so small that 
$\delta\psi$ does not reach the values of order one 
during the time period
$s \in [-\pi/2;\, 0]$. If the particle 
is still in the vicinity of the point $\psi_0$ at $s = -\pi/2$, 
it remains for sure in this vicinity at $s \in [-\pi;\; -\pi/2]$,
because the r.h.s. of Eq. \eqref{psi} is positive again. In 
this way one obtains solutions, which spend two, three, etc.  
sphaleron periods at $\psi \approx \psi_0$ before escaping into the
asymptotic regions $\psi \to \pm\infty$. In the 
leading 
order in $b$ all these solutions correspond to the identical
initial state, 
and (in the case of classically forbidden transitions) 
to the same value of the suppression exponent. 
However, 
an accurate study of the dynamics in the vicinity of the the sphaleron
is generically required  
to obtain the correct value of the suppression exponent in the case 
$b\sim 1$, cf. Ref.~\cite{Levkov:2007ce}.

\end{document}